\documentclass[journal=jacsat,manuscript=article]{achemso}
\pdfoutput=1

\usepackage{amsmath, amssymb}

\usepackage{placeins}

\newcommand{\dd}{\mathrm{d}}

\author{Michiel Laleman}
\affiliation[]{Institute for Theoretical Physics, KU Leuven, Celestijnenlaan 200D, Leuven, Belgium}
\author{Marco Baiesi}
\affiliation{Department of Physics and Astronomy, University of Padova, Via Marzolo 8, Padova, Italy}
\alsoaffiliation{INFN, Sezione di Padova, Via Marzolo 8, Padova, Italy}
\author{Boris P. Belotserkovskii}
\affiliation{Department of Biology, Stanford University, 371 Serra Mall, 
Herrin Labs, Stanford, California 94305-5020, USA}
\author{Takahiro Sakaue}
\affiliation[]{Department of Physics, Kyushu University, Fukuoka 819-0395, Japan}
\author{Jean-Charles Walter}
\affiliation{Laboratoire Charles Coulomb UMR 5221, Universit\'e de Montpellier \& CNRS,  
F-34095, Montpellier, France}
\alsoaffiliation{Laboratoire de Microbiologie et G\'en\'etique Mol\'eculaires UMR 5100,
CNRS \& Universit\'e Paul Sabatier, F-31000 Toulouse, France}
\author{Enrico Carlon}
\affiliation{Institute for Theoretical Physics, KULeuven, Celestijnenlaan 200D, Leuven, Belgium}

\title{Torque-induced rotational dynamics in polymers: Torsional blobs and thinning}


\begin{document}

\begin{abstract}
By using the blob theory and computer simulations, we investigate the 
properties of a linear polymer performing a stationary rotational motion 
around a long impenetrable rod. In particular, in the simulations
the rotation is induced by a torque applied to the end of the polymer 
that is tethered to the rod.
Three different regimes are found, in close analogy with the case of polymers 
pulled by a constant force at one end. For low torques the polymer rotates
maintaining its equilibrium conformation. At intermediate torques the
polymer assumes a trumpet shape, being composed by blobs of increasing
size. At even larger torques the polymer is partially wrapped around
the rod. We derive several scaling relations between various quantities
as angular velocity, elongation and torque. The analytical predictions
match the simulation data well. Interestingly, we find a ``thinning"
regime where the torque has a very weak (logarithmic) dependence on the
angular velocity. We discuss the origin of this behavior, which has no
counterpart in polymers pulled by an applied force.
\end{abstract}

\section{Introduction}

Understanding the dynamics of polymers at the nanoscale is a great
challenge, which is not only interesting from the conceptual viewpoint,
but also because it is useful for applications in nanotechnology. The
challenge stems from the fact that even single polymers can display a
complex dynamical behavior. A popular example is translocation through
a nanopore, a process which has attracted quite some attention in the
past years (see Panja et al. and Palyulin et al.~\cite{panj13,paly14}
for a review). In polymer translocation, the molecule is set into motion
by an applied field at the pore side which acts as a linear pulling force.

\begin{figure*}[t]
\includegraphics[width=0.95\textwidth]{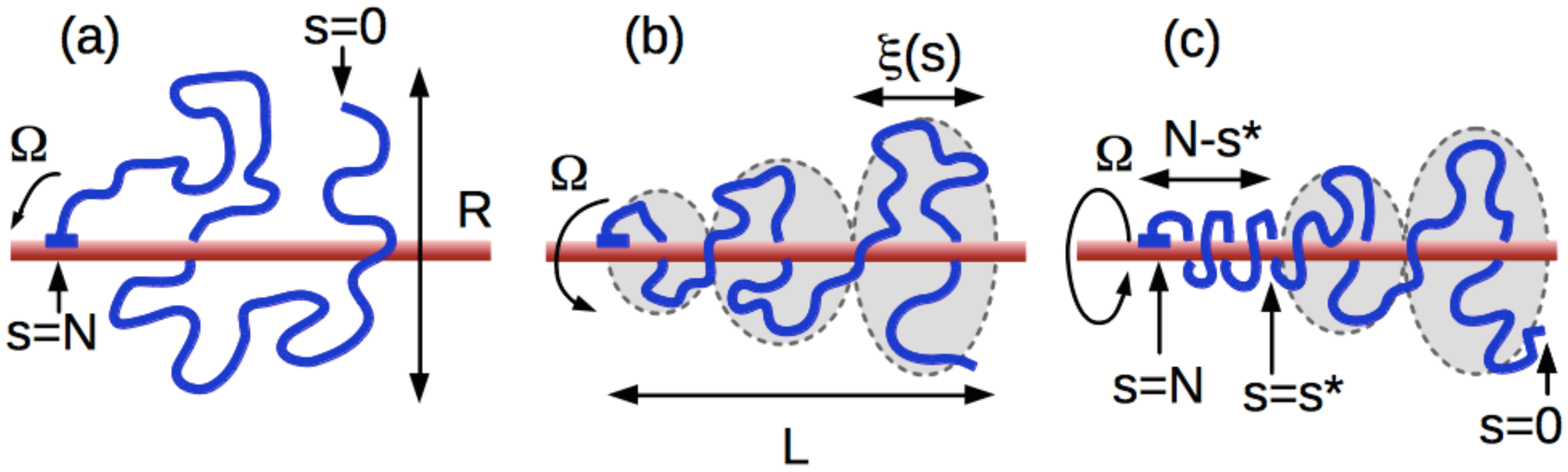}
\caption{Various dynamical regimes for the steady state motion of a
flexible polymer rotating with constant angular velocity $\Omega$ around a
rod. (a) The equilibrium regime (small $\Omega$). (b) The trumpet regime
(intermediate $\Omega$). (c) The stem-trumpet regime (large $\Omega$).
The polymer is attached to the rod at one end $s=N$ while the opposite
end ($s=0$) is free, where $s$ labels the monomers along the chain.
In the equilibrium regime, the shape is on average spherical with a size
$R\simeq a N^\nu$. In the trumpet and stem-trumpet regime the polymer
conformation is characterized by a series of blob of size $\xi(s)$,
which decreases starting from the free end ($s=0$).}
\label{fig:1}
\end{figure*}

While the dynamics of pulled polymers has been thoroughly studied, less
attention has been devoted to the properties of polymers performing
rotational motion, which can be induced by applying a torque. There
are several examples in which the rotational motion of a polymer
is of relevance. For instance, in the DNA double helix melting the
two strands have to unwind from each other to separate. Another
example is the transcription process, where RNA polymerase performs
a rotational motion along the DNA axis wrapping the emergent mRNA
molecule along the DNA~\cite{Belotserkovskii2011,belotserkovskii14}.
The torque caused by relative rotation of the nascent RNA and DNA during
transcription induces dynamic supercoiling in the transcribed DNA (``twin
supercoiled domains'') which have a number of biological implications
\cite{liu87,tsao89}. Also the closure of DNA bubbles involves rotational
dynamics~\cite{dasanna12,dasanna13} and rotational friction is involved
in supercoil formation in DNA~\cite{wada09}.

The aim of this paper is to investigate the torque-induced dynamics on
a flexible polymer using a theoretical model based on torque-balance
equations and computer simulations. We find a rich dynamical behavior
where the polymer conformation is characterized by three different
regimes, illustrated in~\ref{fig:1}. In general the theoretical
predictions are in very good agreement with the simulation results
and there are many analogies with the steady state motion of a polymer
pulled by a force. In addition, the rotational motion is characterized
by a ``thinning'' regime, which has no analogous counterpart in 
the dynamics of pulled polymers.

An isolated polymer in equilibrium is characterized by a single
length scale, its Flory radius $R \simeq a N^\nu$ (where $N$ is the
number of monomers and $a$ the monomer-monomer distance). In a polymer
subject to confinement or to some external forces, instead, new length
scales appear and the aforementioned global scaling generally breaks
down. In these cases, the polymer can be described as being composed
of blobs~\cite{degennes} of characteristic size $\xi$. Depending on the
system considered, the blobs can have constant size or their size may vary
along the chain. According to the blob picture, the equilibrium scaling
governed by the Flory exponent $\nu$, still holds at distances below
$\xi$, while the effect of the external perturbation becomes apparent
at higher distances.  A classical illustration of the blob model is
a polymer in equilibrium stretched by forces applied at the two end
monomers and pointing in opposite direction~\cite{pincus76}. Here blobs
have a characteristic homogeneous size $\xi \simeq k_B T/f$, where $f$
is the applied force.

The blob model has been applied to many other equilibrium
problems in polymer physics, for instance in modeling dense
melts~\cite{degennes}, polyelectrolytes stretched by an applied
force~\cite{saleh2009, stevens2013}, polymers and biopolymers
confined in nano-channels~\cite{nanoslit,reisner05,saka06}
and polymers adsorbed to surfaces~\cite{adsorbed}.  The
blob model also provides a powerful tool to study polymer
dynamics, a paradigmatic example being a polymer pulled by one
end~\cite{brochard93,brochard94,Perkins1995,obermayer07,sakaue12,rowg12}
(briefly reviewed in the following section).  Other out-of-equilibrium
cases where the blob model has been applied include the already
mentioned polymer translocation~\cite{sakaue07,rowg12,sakaue12},
the DNA hairpin folding~\cite{frederickx14}, the DNA compression in a
nanochannel~\cite{khor14} and the unwinding relaxation of a polymer from
a fixed axis~\cite{walter14,walter14b,walter13,belotserkovskii14}.

This paper discusses the steady state dynamics of a flexible polymer
performing a rotational motion around a fixed and impenetrable
rod~\cite{belotserkovskii14,walter14b}.  This motion is induced by a
constant torque applied at the one end monomer that is attached to the
rod.  Hence it is different from the case in which the torque is applied
to induce an internal twist dynamics on the polymer~\cite{wada09}. We
consider polymers containing simple single covalent bonds without
twisting energy.

After an initial transient regime the polymer performs a
steady rotation with a constant angular velocity $\Omega$, which is
the rotational counterpart of the steady linear velocity of a polymer
pulled by a constant force.  To explain the steady state dynamics, we
develop a blob model in which {\it torsional} blobs have a rotational
origin. They have different scaling properties from the {\it tensile}
blobs arising in stretched or dragged polymers.  Our theory shows
that, depending on the torque (or angular velocity), the polymer
can assume different conformations.  For low torques one finds an
equilibrium regime (\ref{fig:1}(a)), followed by a trumpet regime at
intermediate torques (\ref{fig:1}(b)) and a stem-trumpet regime at
high $\Omega$ (\ref{fig:1}(c)).  Because of these regimes, the torque
grows monotonically but not always linearly with the angular velocity.
The theory is supported by Langevin dynamics computer simulations.

\section{Theory}

It is convenient to recall a few basic properties of polymers
pulled at constant force, a case which has been well studied in the
past~\cite{brochard93,brochard94,buguin96,brochard95,obermayer07,rowg12,sakaue12}.
This will allow us to establish some useful scaling relations, which
will be compared to those for polymers under torque. The comparison will
be made through the introduction of torsional~\cite{belotserkovskii14}
blobs. Using the blob model we derive the relation between torque and
rotational velocity. Finally the elongation of a self-avoiding polymer
along the rod is investigated. Part of the theory of a stationary rotating
polymer has already been developed in Ref.~\cite{belotserkovskii14}. Here
we extend this theory to self-avoiding polymers and compute new quantities
as the polymer elongation along the rod, which was not considered
previously.


\subsection{Tensile blobs}

Consider a polymer, consisting of $N$ monomers, pulled by a constant
force at one end.  To label the monomers along the chain a coordinate $s$
is used: the polymer is pulled at $s=N$, while $s=0$ is the free end.
After a transient phase, the polymer arrives in the steady state, moving
with a constant velocity ${\mathrm v}$.  The force difference between two
closely lying points $s$ and $s+\mathrm{d}s$ is then fully compensated
by the friction.  The force balance yields
\begin{eqnarray}
\dd f = \gamma_0 {\mathrm v} \dd s,
\label{dfds_tensile}
\end{eqnarray}
where $\gamma_0$ is the friction per monomer. Note that we focus here to
the Rouse dynamics case, to which the simple relation \eqref{dfds_tensile}
applies. With hydrodynamic interactions, the effect of the conformation
dependent friction should be taken into account \cite{sakaue12}. The
hydrodynamic case is briefly discussed at the end of this Section.

The differential equation \eqref{dfds_tensile} has solution $f(s)=
\gamma_0 {\mathrm v} s$, where we used the boundary condition $f(0)=0$.
We obtain the following relation between the pulling force $f=f(N)$
and the velocity:
\begin{eqnarray}
f = \gamma_0 {\mathrm v} N.
\label{fvN}
\end{eqnarray}
The force is related to the blob size as
\begin{eqnarray}
\frac{\xi(s)}{a} \simeq \frac{k_B T}{f(s) a} \simeq 
\frac{k_B T}{a \gamma_0 {\mathrm v} s}
\simeq \frac{a}{\tau_0 {\mathrm v} s},
\label{xi_tensile}
\end{eqnarray}
where we introduced a monomer-scale characteristic time
\begin{eqnarray}
\tau_0 = \frac{\gamma_0 a^2}{k_B T}.
\label{def_tau0}
\end{eqnarray}
(throughout the paper we use the symbol $\simeq$ to indicate a
``quasi-equality" between two quantities, meaning that numerical
prefactors of the order unity are neglected).  Weak forces do not
significantly perturb the shape of the polymer.  The threshold force
which produces a perturbation can be obtained from the requirement that
the size of the initial blob, $\xi(N)$, exceeds the equilibrium radius
of the polymer: $\xi(N) \gtrsim a N^\nu$. This leads to
\begin{eqnarray}
f \lesssim \frac{k_B T}{aN^\nu}.
\label{feq}
\end{eqnarray}
For stronger forces the polymer consists of blobs of varying size and
distributed from the free end according to \eqref{xi_tensile}, i.e. the
blob size grows towards the free end.  This so-called trumpet regime
ends when the smallest blob has a size comparable to the monomer-monomer
distance, i.e. $\xi(N) \simeq a$.  The trumpet regime appears for forces
in the following range:
\begin{eqnarray}
\frac{k_B T}{aN^\nu} \lesssim  f \lesssim \frac{k_B T}{a} \,.
\label{ftrumpet}
\end{eqnarray}
At higher forces,
\begin{eqnarray}
f \gtrsim \frac{k_B T}{a},
\label{fstem}
\end{eqnarray}
part of the polymer close to the pulled monomer is stretched,
while the end part still has the trumpet shape. 
For simplicity the boundaries between the three regimes were obtained
using static criteria. We note that they can also be obtained from dynamic
criteria involving the drift velocity of the polymer. For instance the
equilibrium regime can be characterized from the requirement
\begin{eqnarray}
{\mathrm v} \tau_R \lesssim a N^\nu,
\label{vel_eq}
\end{eqnarray}
where $\tau_R \simeq \tau_0 N^{1+2\nu}$ is the longest relaxation time. In
the relation \eqref{vel_eq} we require that the distance covered by
the polymer in a time equal to $\tau_R$ is smaller than its equilibrium
radius, i.e. the polymer moves sufficiently slow so that it equilibrates
while leaving a region of the size of its radius.  It can be easily seen
that \eqref{vel_eq} is identical to \eqref{feq}). Analogous arguments
can be used to derive \eqref{ftrumpet} and \eqref{fstem}).

The three regimes characterized by the inequalities
\eqref{feq}-\eqref{fstem} are the tensile counterparts of those shown
in~\ref{fig:1} for polymers under torque.  For comparison with the
rotating polymer it is useful to compute the end-to-end extension of
a polymer pulled by a force.  In the trumpet regime \eqref{ftrumpet},
where the scaling \eqref{xi_tensile} applies to the whole polymer,
the end-to-end distance is given by
\begin{eqnarray}
L = \int_{0}^N \frac{\xi(s)}{g(s)}  \, \dd s\, \simeq  
\frac{a^2}{\tau_0 {\mathrm v}} \left( \frac{\tau_0 {\mathrm v} N}{a} \right)^{1/\nu},
\label{L_pulled}
\end{eqnarray}
where $g(s) \simeq (\xi/a)^{1/\nu}$ is the number of monomers per blob.
Note that at the threshold of the equilibrium regime $f \simeq k_B T/a
N^\nu$ the force-velocity relation \eqref{fvN} yields:
\begin{eqnarray}
{\mathrm v} \simeq \frac{f}{\gamma_0 N} \simeq \frac{a}{\tau_0 N^{1+\nu}}.
\end{eqnarray}
Substituting this expression in \eqref{L_pulled} we recover the
equilibrium result $L \simeq a N^\nu$.


\subsection{Torsional Blobs}

Consider now the rotational counterpart of a pulled polymer,
where a torque $M$  is applied to the monomer attached to the
rod~\cite{walter14b}.  For weak torques we expect that, analogous to the
case of a pulled polymer, the polymer will perform a rotational motion
while maintaining its equilibrium shape.  Using the threshold value for
the polymer pulled by a constant force \eqref{feq}, an upper bound for
the torque in the equilibrium regime is found:
\begin{eqnarray}
M \simeq f R \lesssim k_B T.
\label{eqM}
\end{eqnarray}
Here $f$ is the characteristic value of the applied force and $R \simeq a
N^\nu$ the equilibrium radius.
Using $\mathrm{v} \simeq \Omega R$, which
relates linear to rotational velocity, together with the force-velocity
expression \eqref{fvN}, the inequality \eqref{eqM} can be rewritten:
\begin{eqnarray}
\frac{M}{k_B T} \simeq 
\frac{\gamma_0 N \Omega R^2}{k_B T}  =
\tau_0 \Omega N^{1+2\nu} \lesssim 1.
\label{M_eq_reg}
\end{eqnarray}
This establishes the threshold value for the equilibrium regime for
given $N$ and $\Omega$ (the characteristic monomer-scale time, $\tau_0$,
is defined in \eqref{def_tau0}). Note the similarities of the previous
inequality with \eqref{vel_eq}: the inequality in \eqref{M_eq_reg}
requires the relaxation of the rotating polymer while performing a
fraction $1/2\pi$ of a turn. It should be stressed that the condition
for the equilibrium regime $M/k_BT \lesssim 1$ \eqref{eqM} is
correct up to logarithmic terms which, because they are dimensionless,
cannot be extracted from scaling considerations. A more accurate handling
of these terms is reported in the Section discussing the dependence of
the torque on $\Omega$.

We now use a torque balance argument to set up a differential equation
for $M(s)$, the torque $M$ as a function of the monomer coordinate
$s$. This is the rotational version of \eqref{dfds_tensile}, which
related the force to $s$ in the tensile case.  Consider a torsional blob
of size $\xi(s)$ which rotates around the rod at fixed angular velocity
$\Omega$. The infinitesimal torque difference is $\dd M \simeq \xi \dd f$,
while the friction $\gamma_0 v \dd s \simeq \gamma_0 \Omega \xi \dd s$. The
force balance then gives
\begin{eqnarray}
\frac{\dd M}{\dd s} \simeq \gamma_0 \Omega \xi^2(s).
\label{force_balance}
\end{eqnarray}
Unlike in the pulling case, \eqref{dfds_tensile}, the blob size $\xi(s)$
appears in the previous equation, reflecting the fact that the rotational
friction is conformation-dependent even without the hydrodynamic
interactions. To close \eqref{force_balance}, we thus need to know the
local relation between the torque and the torsional blob size. To this end
we consider the equilibrium properties of a polymer wound around a rod.

The winding angle $\theta$ of a polymer attached to a rod is
defined as $2\pi$ times the number of turns the polymer performs
around the rod, starting from the fixed end. In equilibrium the
average winding angle vanishes by symmetry: $\langle \theta \rangle
=0$. This variable is distributed according to the following scaling
form~\cite{fisher84,rudnick87,duplantier88,walter11}:
\begin{eqnarray}
P_\theta (\theta, N) = 
f_\theta \left( \frac{\theta}{(\ln N)^\alpha} \right),
\label{Ptheta}
\end{eqnarray}
where $f_\theta$ is a scaling function. This form is generic, but the
case of the two dimensional winding of an ideal polymer around a circle
one has $\alpha=1$~\cite{rudnick87}. For a two dimensional self-avoiding
polymer $\alpha=1/2$~\cite{fisher84,duplantier88}.  In three dimensions
the scaling for the ideal polymer remains $\alpha=1$. The equilibrium
three dimensional self-avoiding polymer wound around the rod has been
studied numerically~\cite{walter11} and data were fitted with $\alpha
\approx 0.75$. In the torque ensemble the probability distribution is
obtained by a Laplace transform of \eqref{Ptheta}:
\begin{eqnarray}
P_M (M, N) 
&=& \int \dd \theta \, e^{\beta M \theta} \,P_\theta (\theta, N) \nonumber\\
&=& f_M\left( \frac{M {(\ln N)^\alpha}}{k_B T} \right).
\end{eqnarray}
with $f_M$ a scaling function. This relation implies that typical values
scale as $M \simeq k_B T/(\ln N)^{\alpha}$.
Applying this result to
the torsional blob $\xi(s) \simeq a g(s)^{\nu}$ with $g(s)$ monomers,
we get the torsional blob-torque relation
\begin{eqnarray} 
M(s) &\simeq &\frac{k_BT}{[\ln{(\xi(s)/a)}]^{\alpha} }
\label{scal_M_xi}
\end{eqnarray}
Differentiating the previous relation with respect of $s$ we find:
\begin{eqnarray}
\frac{\dd M}{\dd s} & \simeq & -\frac{k_BT}{\xi [\ln{(\xi/a)}]^{\alpha+1}} 
\frac{\dd \xi}{\dd s}
\label{diff_M2}
\end{eqnarray}
Combining \eqref{force_balance} and \eqref{diff_M2} we obtain a 
differential equation for $\xi$. This can be solved explicitely 
if one neglects the logarithmic term, which has a very weak 
dependence on $\xi$. The solution is
\begin{eqnarray}
\frac{\xi(s)}{a} \simeq \left( \frac{1}{\tau_0 \Omega s}  \right)^{1/2}.
\label{xi_blob}
\end{eqnarray}
which agrees with the result of Eq.~(46) of
Ref.~\cite{belotserkovskii14}. Note the central difference with respect
to the tensile blobs in \eqref{xi_tensile}, where the decay in $s$
is much faster: $\xi \sim 1/s$.

From (\ref{xi_blob}) we get the threshold for the different regimes in
the same way as for tensile blobs. Again, the equilibrium regime ends
when the equilibrium polymer radius is equal to the size of the initial
blob $\xi(N)$, i.e. $\xi(N) \simeq a N^\nu$.  From this relation, and
using \eqref{xi_blob}, we get again \eqref{M_eq_reg}.  The boundary
between the trumpet and stem-trumpet regime is obtained when $\xi(N)
\simeq a$, which yields
\begin{eqnarray}
\tau_0 \Omega N \simeq 1\,.
\label{onset_stem}
\end{eqnarray}

In summary: we expect three distinct regimes, with boundaries given by
the analysis above.  These regimes are also depicted in~\ref{fig:1}.
\begin{itemize}
\item[(a)] If $\tau_0 \Omega \lesssim 1/N^{1+2\nu}$, the polymer rotates
while maintaining its {\it equilibrium shape}.
\item[(b)] If $1/N^{1+2\nu} \lesssim \tau_0 \Omega \lesssim 1/N$ the
polymer assumes a {\it trumpet shape} in its full length. It is composed
of torsional blobs of decreasing size $\xi \sim s^{-1/2}$ starting from
the free polymer end $s=0$.
\item[(c)] If $\tau_0 \Omega \gtrsim 1/N$ the polymer assumes a trumpet
shape in a limited range $0 \leq s \leq s^*$, while the monomers close
to the tethered point, $s^* \leq s \leq N$, are fully wrapped around
the rod. This marks the {\it stem-trumpet} regime. $s^*$, the boundary
between stem and trumpet, can be obtained from the relation $\xi(s^*)/a
\simeq 1$, thus:
\begin{eqnarray}
s^* \simeq \frac{1}{\tau_0 \Omega} \,.
\label{def_s*}
\end{eqnarray}
\end{itemize}
Note that the size of the blob diverges as $s \to 0$.  To avoid
divergences in the following calculations we compute $\xi_0$, the size
of the `last' blob, containing $s_0$ monomers. The size can be obtained
using the blob scaling \eqref{xi_blob} (recall that $s=0$ is the free end)
and the equilibrium properties of the last blob ($\xi_0 \sim a s_0^\nu$):
\begin{eqnarray}
\frac{\xi_0}{a} \simeq \left( \frac{1}{\tau_0 \Omega}  \right)^{1/2} 
\left(\frac{\xi_0}{a}\right)^{-1/2 \nu}.
\end{eqnarray}
Rewriting yields expressions for $\xi_0$ and $s_0$:
\begin{eqnarray}
\frac{\xi_0}{a} \simeq 
\left( \frac{1}{\tau_0 \Omega}  \right)^{\nu/(1 + 2 \nu)} ,
\label{last_xi}
\end{eqnarray}
\begin{eqnarray}
s_0 \simeq  \left( \frac{1}{\tau_0 \Omega}  \right)^{1/(1 + 2 \nu)}.
\label{last_s}
\end{eqnarray}
At the threshold between equilibrium and trumpet regime $\tau_0 \Omega
N^{1+2\nu} \simeq 1$ and from \eqref{last_xi} one then finds $\xi_0 \simeq
a N^\nu$, i.e. the last blob `covers' the whole polymer.


\subsection{Torque vs. angular velocity}

The analysis of the previous section can be used to compute the total
torque applied as a function of the angular velocity. The fact that the
rotational friction depends on the conformation leads to a non-trivial
dependence of $M$ on $\Omega$.

\subsubsection{Equilibrium Regime}

To obtain the torque we integrate \eqref{force_balance}
\begin{eqnarray}
M \simeq \int_{0}^N \dd s \, \gamma_0 \Omega \xi^2(s)
\label{M_int}
\end{eqnarray}
When the whole polymer is in equilibrium, random coil statistics holds:
$\xi^2 (s) \simeq a^2 (N-s)^{2\nu}$. Plugging this into 
\eqref{M_int} we get
\begin{eqnarray}
\frac{M}{k_B T} \simeq \tau_0 \Omega  N^{1+2\nu},
\label{M_eq}
\end{eqnarray}
This calculation reproduces \eqref{M_eq_reg}, which was previously
obtained from scaling arguments.

\subsubsection{Trumpet Regime}

To calculate the torque in the trumpet regime we integrate
Eq.~\eqref{diff_M2}.  We use $\xi$ as integration variable, hence the
interval $0 \leq s \leq N$ corresponds to $\xi(N) \leq \xi \leq +\infty$,
as the last blob size formally diverges in the limit $s \to 0$. Here
$\xi(N)$ denotes the size of the first blob.  As \eqref{diff_M2} was
obtained from the differentiating \eqref{scal_M_xi} we get:
\begin{eqnarray}
M = \int_0^N \frac{\dd M}{\dd s} \, \dd s \simeq \frac{k_BT}{\left[ \ln (\xi(N)/a)\right]^\alpha}
\label{M_trump0}
\end{eqnarray}
Using \eqref{xi_blob} we obtain:
\begin{eqnarray}
\frac{M}{k_BT} \simeq 
\frac{1}{\left[\displaystyle\ln \left( \frac{1}{\sqrt{\tau_0 \Omega N}} \right) \right]^{\alpha}}
\label{M_trump}
\end{eqnarray}
which is a very weakly increasing function of $\Omega$.

We have used Eq.~\eqref{eqM} to estimate the boundary between the
equilibrium and trumpet regimes. Alternatively we can estimate
this boundary by equating the torques of
\eqref{M_eq} and \eqref{M_trump0} and assuming that the polymer consists
of a single blob $\xi(N) \simeq a N^\nu$. We obtain
\begin{eqnarray}
\tau_0 \Omega N^{1+2\nu} \simeq \frac{1}{\left[ \ln N^\nu \right]^\alpha}
\label{new_bound}
\end{eqnarray}
hence the rotating polymer is in
equilibrium for torques 
\begin{eqnarray}
\frac{M}{k_BT} \lesssim \frac{1}{\left[ \ln N \right]^\alpha}
\end{eqnarray}

Compared to \eqref{M_eq_reg} the previous relation contains a logarithmic
term. In practice this term does not affect
strongly the estimate of the boundary of the equilibrium regime.
However the estimate \eqref{new_bound} is more accurate as it can be seen from a
following argument. We can estimate the boundary using
\begin{eqnarray}
M \theta \simeq k_B T
\label{M_theta}
\end{eqnarray}
where $\theta$ is the stationary winding angle of the rotating polymer.
Eq.~\eqref{M_theta} is the rotational counterpart of the relation $f R
\simeq k_B T$ (see \eqref{feq}) which is an estimate of the boundary of
the equilibrium regime for a polymer pulled by a force $f$ on one end
($R \simeq a N^\nu$ is the Flory radius). In the rotating
case the polymer does not display a substantial deformation compared to
equilibrium when the average winding is smaller than that from equilibrium
fluctuations in absence of torque. The boundary of the equilibrium regime
can be estimated by equating the average winding angle induced by the
torque to the equilibrium value in absence of torque as follows:
\begin{eqnarray}
\theta \simeq \sqrt{\langle \theta^2 \rangle_{M=0}}
\label{eq_fluct_theta}
\end{eqnarray}
Now, $\theta$ is obtained from \eqref{M_theta}, using the torque for
the equilibrium regime \eqref{M_eq}.  As equilibrium fluctuations are
described by the scaling form \eqref{Ptheta} we get $\langle \theta^2
\rangle_{M=0} \sim (\ln N)^{2\alpha}$. Combining these results we get
again the relation \eqref{new_bound}.

\begin{figure}[t]
\begin{center}
\includegraphics[width=0.8\textwidth]{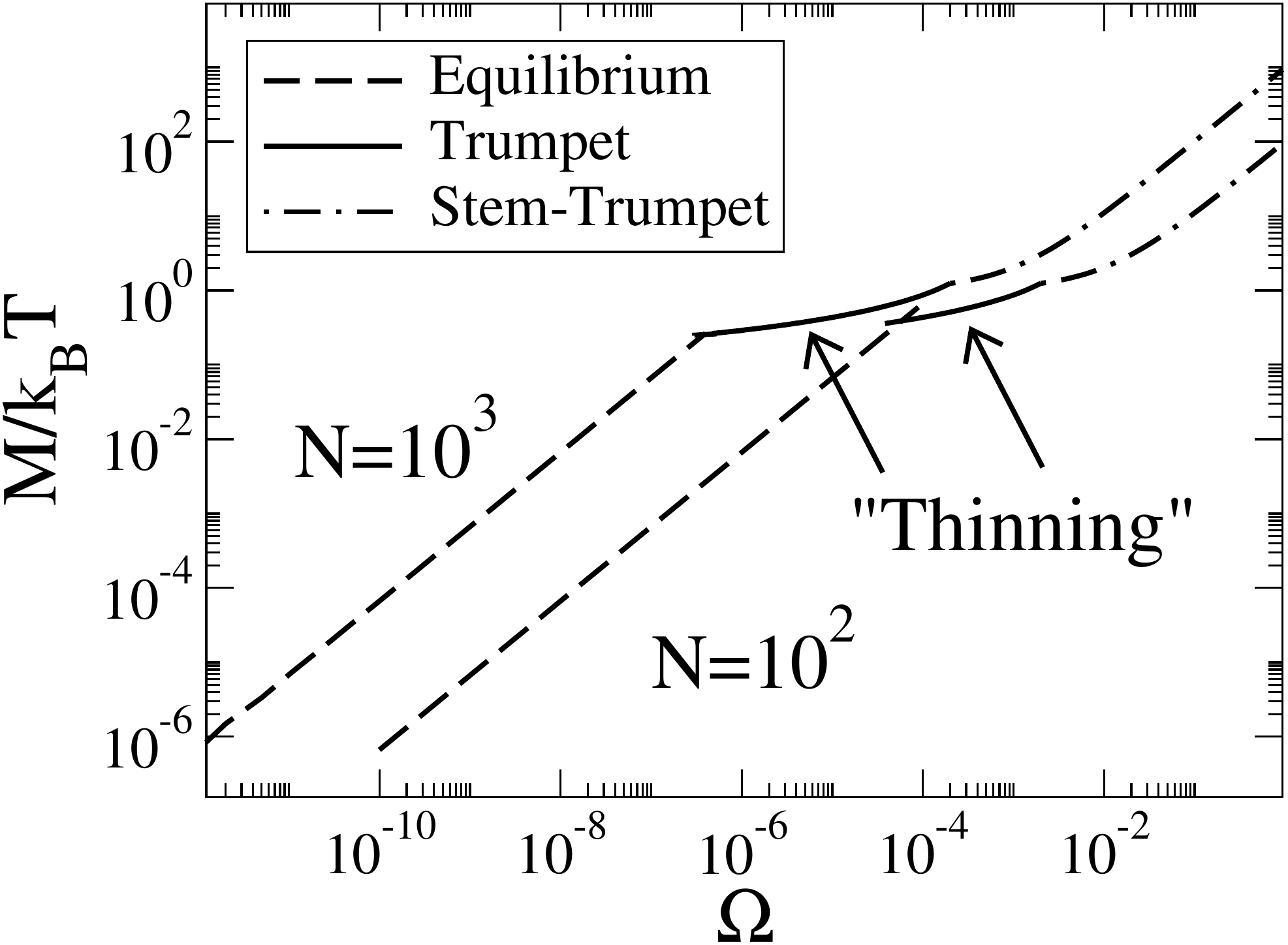}
\caption{Plots of torque vs. angular velocity predicted by the theory
(\eqref{M_eq}, \eqref{M_trump} and \eqref{M_stem}), for ideal polymers of
two different lengths ($N=10^2$ and  $N=10^3$).  The three regimes are
highlighted. In the equilibrium regime and in the stem trumpet regime
$M$ is linear in $\Omega$ (in the latter asymptotically in $N$). In
the intermediate trumpet regime $M$ depends weakly (logarithmically)
on $\Omega$.  A cutoff value $\Delta = 0.2$ was used as estimate the
boundary between the trumpet and stem-trumpet regimes.}
\label{fig:MvsOmega}
\end{center}
\end{figure}

\subsubsection{Stem-Trumpet Regime}

In the stem-trumpet regime the polymer is assumed to form a tight helix
for $s^* \leq s \leq N$, where $s^*$ is given by \eqref{def_s*} and a
trumpet for $s \leq s^*$. The contribution of the torque from the stem
can be obtained by integrating \eqref{xi_blob} where the blob size fixed
to $\xi \simeq R_r$, with $R_r$ the radius of the rod. One finds $M_{\rm
stem} \simeq \gamma_0 \Omega R_r^2 (N-s^*)$. The trumpet part contributes
to the torque as \eqref{M_trump}, with $s^*$ replacing $N$. Combining
these two contributions results in:
\begin{eqnarray}
\frac{M}{k_B T}\simeq\left( \frac{R_r}{a} \right)^2 \left( N - s^* \right) \tau_0 \Omega
+  \frac{1}{\left[ \displaystyle \ln \left( \frac{1}{\sqrt{\tau_0 \Omega s^*}} 
\right) \right]^{\alpha}}
\label{M_stem}
\end{eqnarray}
In this regime the torque is again, to leading order, proportional to
$\Omega$.  Eq. \eqref{onset_stem} provides an estimate of the boundary
between the trumpet and the stem-trumpet regimes. If this relation is
used as a strict equality $\tau_0 \Omega N =1$ one gets a divergence
of \eqref{M_trump}.  Analogously, \eqref{M_stem} is divergent when
\eqref{def_s*} is used as a strict equality. In order to avoid such
effects we used in both cases a different cutoff $\tau_0 \Omega N =\Delta$
and $\tau_0 \Omega s^* = \Delta$, where $\Delta < 1$.  The torque in
the trumpet regime is therefore bounded in the interval:
\begin{eqnarray}
\frac{1}{(\ln N)^\alpha} \lesssim \frac{M}{k_B T} \lesssim 1
\end{eqnarray}

In summary: the torque-angular velocity relation is highly non-trivial,
as depicted in~\ref{fig:MvsOmega}. At weak and high torques, respectively
the equilibrium and stem-trumpet regimes, $M$ is proportional to $\Omega$.
This can be directly seen in \eqref{M_eq} and \eqref{M_stem} (in the
latter case this proportionality holds in leading order for large $N$).
Both regimes are characterized by two different rotational friction
coefficients $M_{\rm C} \simeq \Gamma_{\rm C} \Omega$ and $M_{\rm H}
\simeq \Gamma_{\rm H} \Omega$, with different dependencies on the
polymer length: $\Gamma_{\rm C} \propto N^{1+2\nu}$ and $\Gamma_{\rm
H} \propto N$.  In the intermediate trumpet regime there is a strong
``thinning'' behavior with $M$ increasing very weakly with $\Omega$.
Here the polymer decreases its resistance to the rotational motion by
getting closer to the rod axis.  This contrasts the stretching by a
linear force, where the force-velocity relation is always linear for
Rouse dynamics, as can be seen by inspection of the force-velocity
relation \eqref{fvN}. Only for chains with hydrodynamic interactions,
a non-linear force-velocity relation is obtained \cite{sakaue12}.


\subsection{Polymer elongation}

A self-avoiding polymer is expected to, at sufficiently high torque,
elongate as a helix along the rod. The blob theory developed in
the previous section can be used to compute the elongation in the
three regimes. The elongation, denoted by $L$, is defined as the
distance between the two end monomers along the rod direction. In the
calculations we can safely use the leading order form for the blob size
\eqref{xi_blob}, ignoring higher order logarithmic terms.

\subsubsection{Equilibrium regime}

In the equilibrium
regime the polymer maintains its equilibrium shape, hence
\begin{eqnarray}
L \simeq a N^\nu.
\label{Leq}
\end{eqnarray}

\subsubsection{Trumpet regime}

In the trumpet regime one can simply sum up the contribution of the
different blobs:
\begin{eqnarray}
L = \int_{0}^N \frac{\xi(s)}{g(s)}  \mathrm{d}s,
\label{eq:integral_elongation}
\end{eqnarray}
where $g(s)$ is the number of monomers in the blob. In the calculation
there is no need to separate the contribution of the last blob
as the singularity in $s\to 0$ is integrable. Using $g(s) \simeq
[\xi(s)/a]^{1/\nu}$ and \eqref{xi_blob}, we can evaluate the integral:
\begin{eqnarray}
L \simeq a \int_{0}^N \left(\frac{1}{\tau_0 \Omega s}\right)^{\frac{\nu-1}{2\nu}}  
\mathrm{d}s\
\simeq \frac{a}{\tau_0 \Omega} \left( \tau_0 \Omega N\right)^{\frac{1+\nu}{2\nu}}.
\label{Ltrumpet}
\end{eqnarray}
This is the rotational counterpart of \eqref{L_pulled}, which was
derived for pulled polymers. The boundary between the equilibrium and
the trumpet regime is characterized by the relation $\tau_0 \Omega \simeq
1/N^{1+2\nu}$. Substituting this boundary condition in \eqref{Ltrumpet}
gives $L \simeq a N^\nu$, as expected in the equilibrium regime. Hence
\eqref{Ltrumpet} merges with \eqref{Leq} at the phase boundary.

\subsubsection{Stem-Trumpet Regime}

In the stem-trumpet regime the first $N-s^*$ monomers, counting from
the end attached to the rod, are fully wrapped around the rod, while the
last $s^*$ are in a trumpet conformation. Assuming that in the stem the
elongation per monomer is $h$, we find
\begin{eqnarray} 
L &\simeq& h (N-s^*) + a \int_{0}^{s^*}
\left(\frac{1}{\tau_0 \Omega s}\right)^{\frac{\nu-1}{2\nu}} \mathrm{d}s\ 
\nonumber \\
&\simeq&
hN + \frac{a-h}{\tau_0 \Omega},
\label{Lstem}
\end{eqnarray}
where we have used \eqref{def_s*}: $s^* \simeq 1/(\tau_0 \Omega)$.
To show the compatibility of \eqref{Ltrumpet} and \eqref{Lstem}, we plug
in $\tau_0 \Omega = 1/N$, which marks the phase boundary between trumpet
and stem-trumpet regime. It is clear that both expressions then coincide
at $L \simeq aN$.

In summary: when the polymer is performing an equilibrium rotation, the
elongation along the rod axis is dictated by the random coil value. Upon
entering the trumpet regime, blobs start to form and $L$ grows as $\sim
\Omega^{(1-\nu)/2\nu}$. Finally, in the stem-trumpet regime the polymer
is wrapped around the rod and asymptotically $L \sim N$. Note that in
an ideal chain monomers can pass through each other, thus there is no
helix formation and the previous calculations are not at issue.


\subsection{Hydrodynamics}

The results presented above can be extended to the
hydrodynamic case. A blob of size $\xi$ moving with velocity $v$
now experiences a friction propotional to its size, $\gamma_0 v \xi /
a$, while without hydrodynamics the friction scales with the number of
monomers in a blob. The contribution of a small segment $\dd s$ of the
polymer to the friction force is then
\begin{eqnarray}
 \dd f = \gamma_0 v \left(\xi / a \right) \frac{\dd s}{\left(\xi / a \right)^{1/\nu} }.
 \label{eq:df_hydro}
\end{eqnarray}
Hence the torque balance ($\dd M =\xi \dd f $) reads
\begin{eqnarray}
\frac{\dd M}{\dd s} \simeq  \gamma_0 a^2 \Omega 
\left( \frac{\xi}{a} \right)^{3-1/\nu},
\label{eq:dM_hydro}
\end{eqnarray}
which is the counterpart of \eqref{force_balance}.

\eqref{diff_M2} was derived from local equilibrium input, hence
it remains valid in the hydrodynamic case. Combining \eqref{diff_M2}
and \eqref{eq:dM_hydro} results in a differential equation 
for $\xi(s)$. The solution, neglecting again logarithmic terms, is 
\begin{eqnarray}
\xi(s) \simeq \left(\frac{1}{\tau_0 \Omega s} 
\right)^{\frac{\nu}{3 \nu - 1}}.
 \label{eq:blobsize_hydro}
\end{eqnarray}
Using the Flory exponent $\nu = 3/5$ we get $\xi \sim s^{-3/4}$. This
is a more rapid decay of the blob size than in the Rouse case $\sim
s^{-1/2}$, see Eq.\eqref{xi_blob}.

Using \eqref{eq:blobsize_hydro} we can calculate the scaling behavior of
various quantities in the hydrodynamic case. Note that \eqref{M_trump0}
implies that $M$ still depends logarithmically on $\Omega$ in
the trumpet regime. 
Another interesting quantity to compute is the elongation in the
trumpet regime:
\begin{eqnarray}
L = \int_{0}^N \frac{\xi(s)}{g(s)}  \mathrm{d}s \simeq 
\frac{a}{\tau_0 \Omega} \left( \tau_0 \Omega N\right)^{\frac{2\nu}{3 \nu - 1}}.
\end{eqnarray}

\begin{figure*}[t]
\begin{center}
\includegraphics[width=0.45\textwidth]{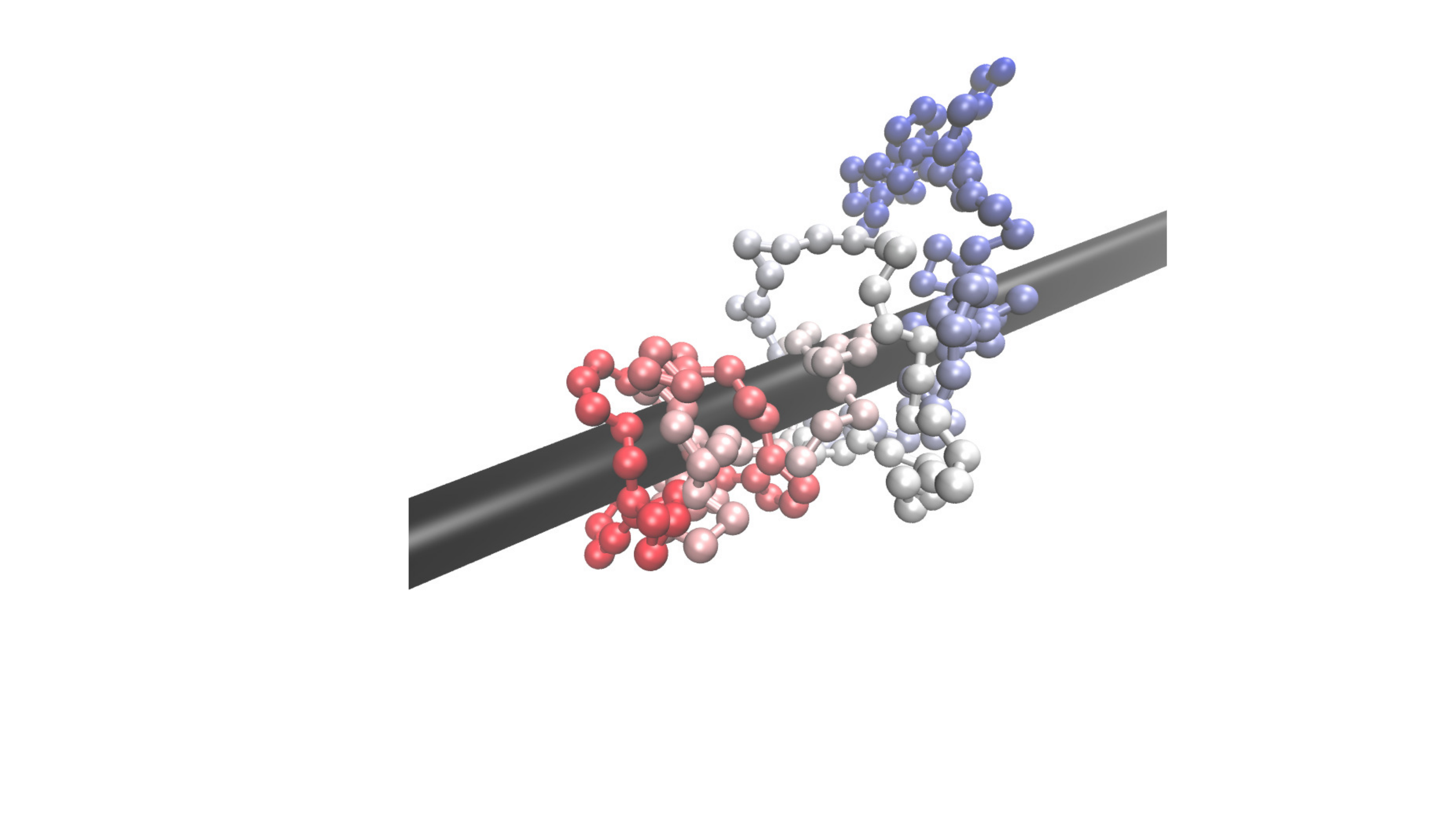}
\includegraphics[width=0.5\textwidth]{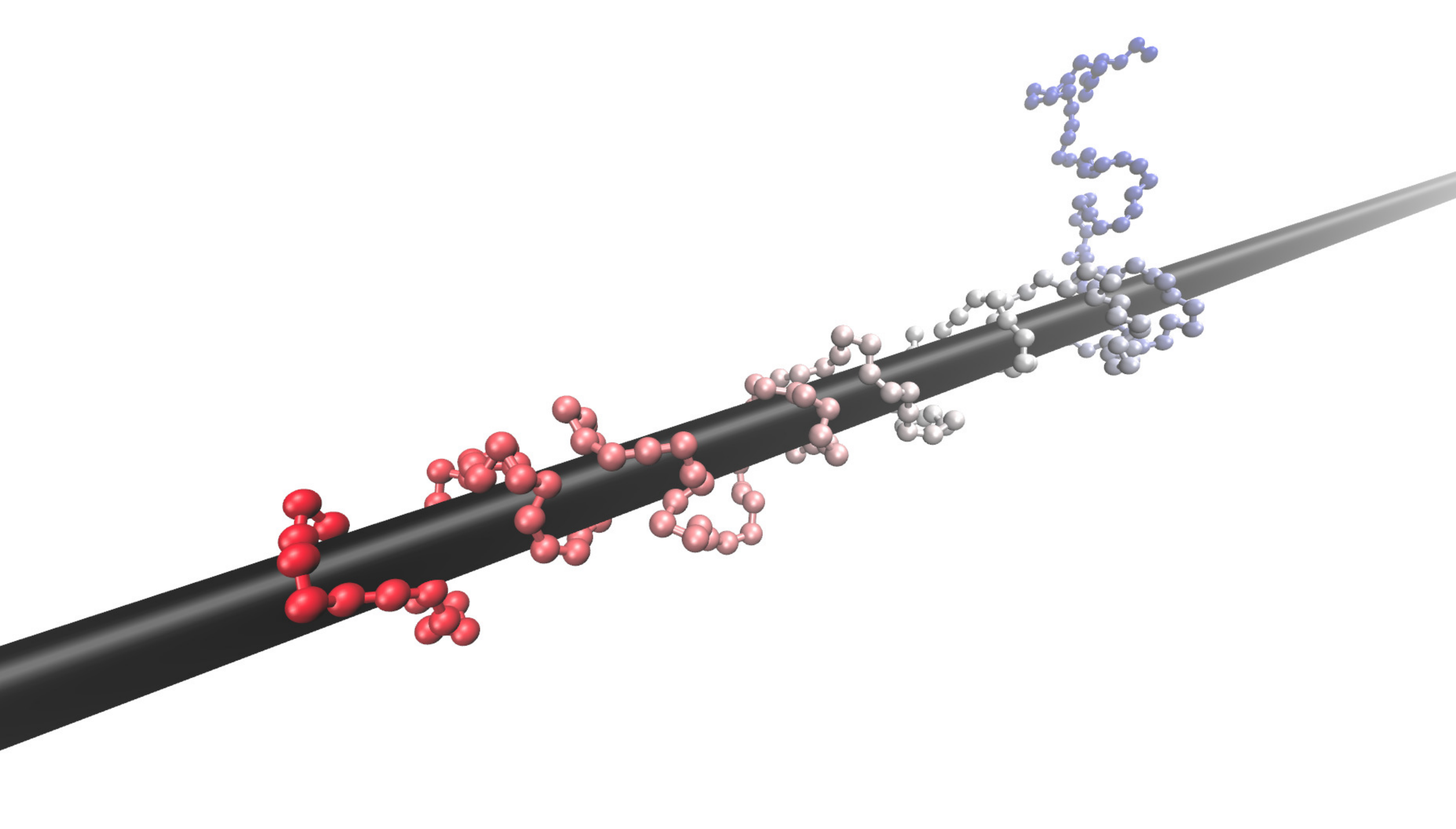}
\caption{Snapshots of simulations of steady rotating polymers around
an infinitely long rod. The figure shows an ideal polymer (left) where
excluded volume interactions are only between each monomer and the
rod and a real polymers (right) where excluded volume interactions act
between all monomers as well. The first monomer in the red end is
attached to the rod and is subjected to an external torque.  The torque
has the value $M=3.55$ and the length is $N=200$ in both cases, which
corresponds to the trumpet regime.}
\label{fig:examples}
\end{center}
\end{figure*}

\FloatBarrier

\section{Numerical Simulations}

Numerical simulations were performed using the LAMMPS package
\cite{lammps}.  Due to the large computational cost of simulating systems
with hydrodynamics interactions, we restrict our analysis to simulations
of Rouse polymers.  The polymer was modeled using a bead spring model with
a finitely extensible nonlinear elastic (FENE) interaction~\cite{FENE}
between successive beads, so that a maximal distance between bonds
is imposed.  The rod was modeled by defining a cylindrical region with
radius $R_r$ around the z-axis and adding soft repulsive interaction
of Weeks-Chandler-Andersen (WCA) type~\cite{WCA}. Because the maximal
distance between beads is chosen smaller than the diameter of the rod,
it is impossible for bonds to cross through the rod.

In order to impose a constant torque we introduced a `ghost' particle at
the origin of the axes, i.e. within the rod. A second particle was kept
at a constant distance from it using the SHAKE algorithm~\cite{shake}.
The distance was set equal to $R_r$, the radius of the rod. A constant
force $f_{b}$ was applied to the second particle, which corresponds to
the first bead of the polymer. This force was oriented perpendicular
to the plane defined by the rod and the axis connecting this first bead
and the ghost particle.  The applied torque equals then $M = f_b R_r$.
Furthermore LJ units of LAMMPS were used, such that $k_B = m = 1$. In
these units the two parameters controlling the repulsive Lennard-Jones
part of the FENE potential (usually called $\epsilon$ and $\sigma$)
are also set to one. Then we chose to set $\gamma_0  =a = R_r =1$. The
maximal bond length, controlled by the FENE potential, was set to 1.5.

The system was integrated using the \textit{fix nve} and \textit{fix
langevin} commands in LAMMPS, which corresponds to performing a Langevin
dynamics simulation.  Time integration is realized using a velocity
Verlet updating scheme.  The coupling to the heat bath is implemented
through the thermostat described in Schneider and Stoll\cite{schneider}
and Dunweg and Paul\cite{dunweg}. Simulations were performed for different
torques and for lengths up to $N=400$.  \ref{fig:examples} shows two
snapshots of the simulations, representing the two models used in the
simulations: ideal polymers (left) and self-avoiding polymers (right).
In the latter case an additional WCA potential was added.

In the simulations the torque is fixed and applied to the anchored
monomer. Thus there are no numerical errors on $M$. The angular velocity
was instead determined from the simulation runs. For every value of the
applied torque the simulation was run for a significantly long time, such
that the stationary state was reached. The angular velocity was determined
by counting the number of turns a monomer made in a given time interval,
multiplying this number by $2\pi$ and dividing by the length of the time
interval. Monomers close to the attached end have larger fluctuations,
so $\Omega$ was estimated by averaging over the end monomers.

In this section we focus on three observables.  The first is the blob
size along the rod.  Due to its interesting properties, a separate
section for the size of the last blob is added.  Then the focus shifts
to the torque versus angular velocity relation.  Finally the elongation
of a SAW along the rod is presented.

\begin{figure*}[t]
\includegraphics[width=0.45\textwidth]{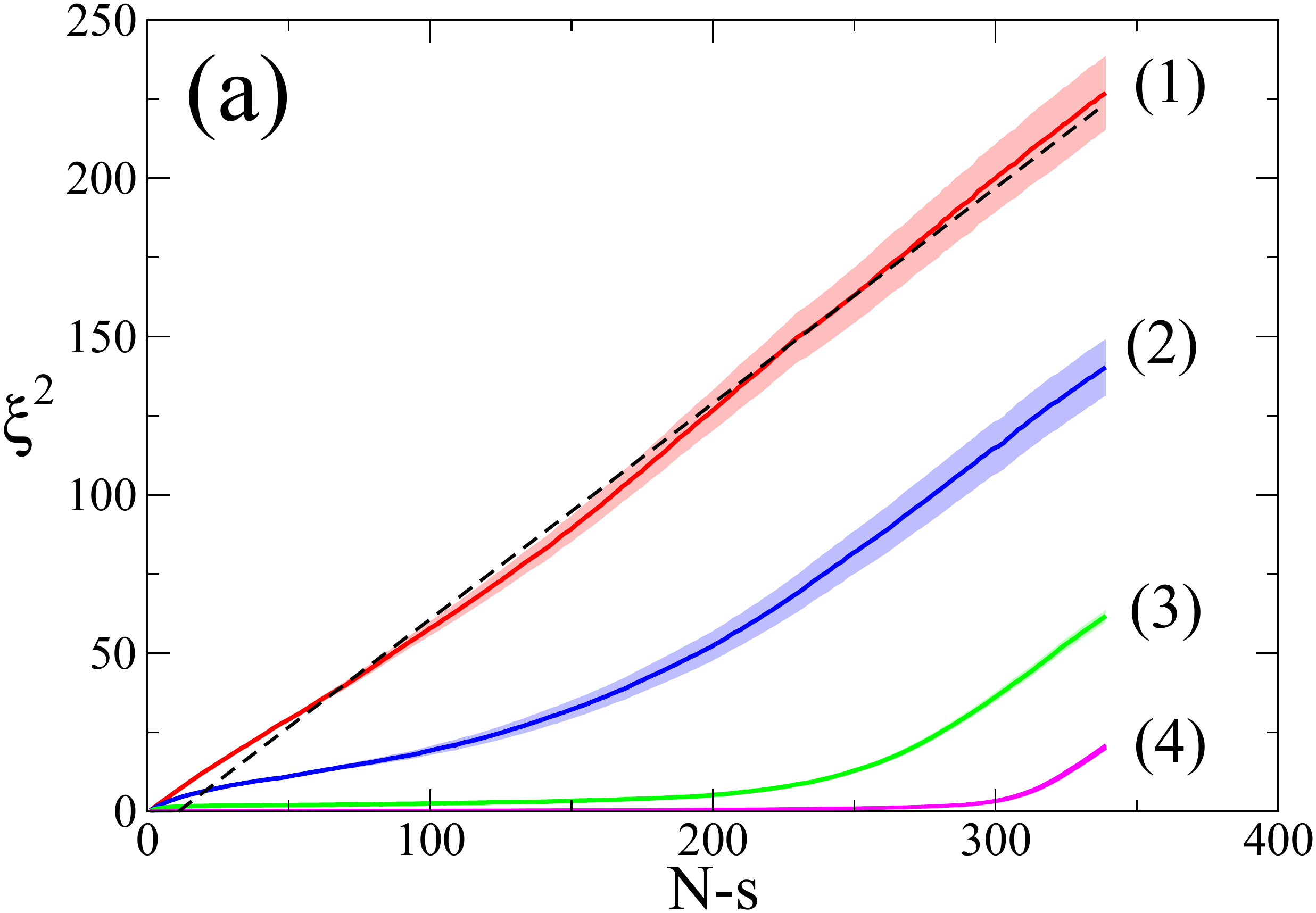}
\includegraphics[width=0.45\textwidth]{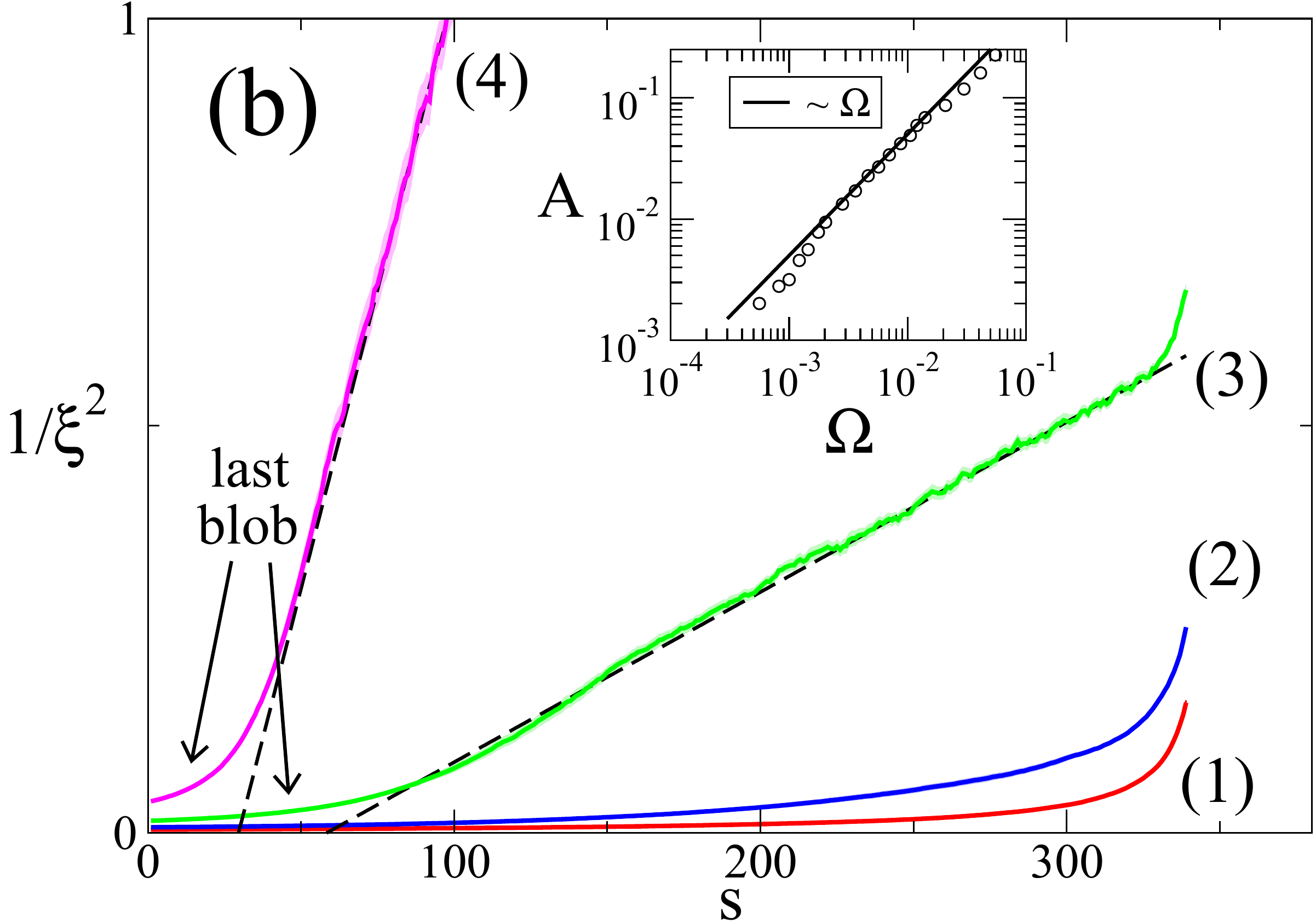}
\caption{Plots of (a) $\xi^2$ vs. $N-s$ and (b) $1/\xi^2$ vs. $s$
for an ideal polymer of length $N=340$ and for three different angular
velocities: (1) $\Omega =5.17 \cdot 10^{-5}$, (2) $\Omega = 1.55 \cdot
10^{-4}$, (3) $\Omega = 5.61 \cdot 10^{-4}$ and (4) $\Omega = 8.19
\cdot 10^{-4}$. The standard error is depicted as a shade around 
the plots.
}
\label{fig:R2vs1overR2}
\end{figure*}

\subsection{Blob size vs. $s$}

First of all we are interested in the blob size.
A good estimator for this blob size is the average distance of each monomer from
the rod. As the rod is oriented along the z-axis we have
\begin{eqnarray}
\xi (s) = \sqrt{x^2(s) + y^2(s)},
\label{defxs}
\end{eqnarray}
where $x(s)$ and $y(s)$ represent the x- and \mbox{y-coordinate} of the
$s$-th monomer.

In~\ref{fig:R2vs1overR2}(a), $\xi^2(s)$ vs. $N\!-\!s$ is presented for
four different values of the applied torque, for ideal polymers. These
four cases are labeled from 1 to 4, ordered according to increasing
applied torque.  We estimated the stationary angular velocity $\Omega$,
for the different cases, as presented in the caption, from the ratio of
the number of turns of each monomer over the simulation time.

For case 1, the weakest applied torque, the polymer has an equilibrium
shape characterized by random coil scaling (recall that $s\!=\!0$ is
the free end):
\begin{eqnarray}
\xi^2 (s) \simeq a^2 (N-s)^{2\nu},
\label{RW_behavior}
\end{eqnarray}
with $\nu = 1/2$ for ideal polymers. In this case we estimate $\tau_0
\Omega N^2 \! \approx \! 6$, using $\tau_0 \! = \! 1$ and setting $\Omega$
as given in the caption of~\ref{fig:R2vs1overR2}.  Since $\tau_0 \Omega
N^{1+2\nu}\! > \!1$, this should correspond to the trumpet regime -
according to the analysis in the theory section. However, the theory
is based on scaling arguments and the relations are valid up to some
numerical prefactors. This first case (1) suggests that the trumpet regime
sets in at $\Omega$'s a bit higher than predicted by \eqref{M_eq_reg}.

When higher torques are applied (cases 2-4) the stationary shape
gets distorted from the equilibrium.  The polymer gets wrapped
around the rod, but close to the free end $s\!=\!0$ the shape is well
fitted by \eqref{RW_behavior}, indicating that the free end is
equilibrated. For case 4 we get $\tau_0 \Omega N\! \approx\! 0.3$,
indicating that the system is close to the onset of the stem-trumpet
regime, see \eqref{onset_stem}.

\begin{figure*}[t!]
\includegraphics[width=0.45\textwidth]{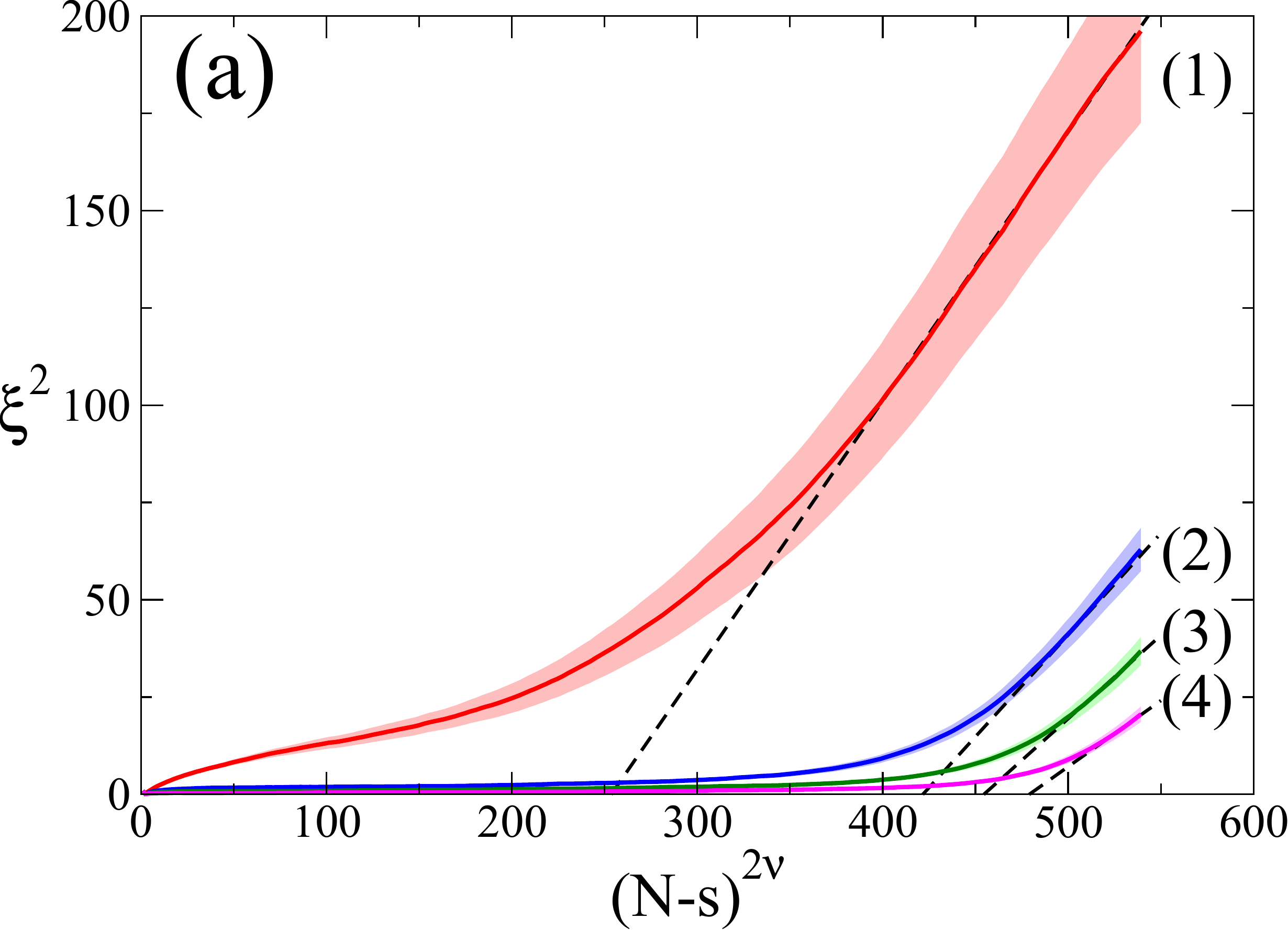}
\includegraphics[width=0.45\textwidth]{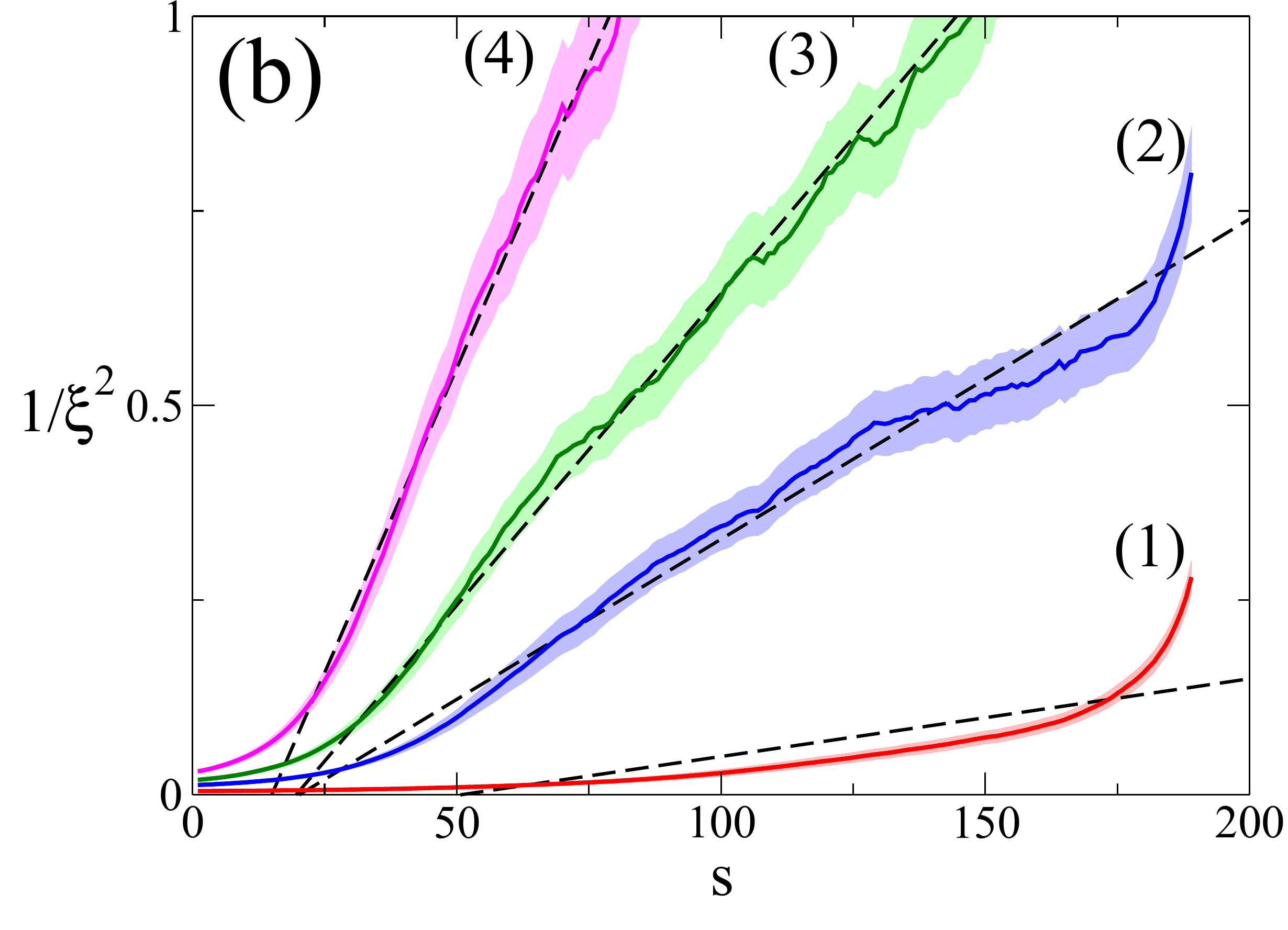}
\caption{Plots of (a) $\xi^2$ vs. $(N-s)^{2\nu}$ ($\nu=0.588$) and (b)
$1/\xi^2$ vs. $s$ for a self-avoiding polymer of length $N=190$ and for
three different angular velocities: (1) $\Omega =1.95 \cdot 10^{-4}$,
(2) $\Omega = 1.16 \cdot 10^{-3}$, (3) $\Omega = 2.34 \cdot 10^{-3}$
and (4) $\Omega = 4.01 \cdot 10^{-3}$.  
The standard error is depicted as a shade around the plots.}
\label{fig:SAW_R2vs1overR2}
\end{figure*}

In order to check the validity of the blob theory we have replotted the
data of \ref{fig:R2vs1overR2}(a) in \ref{fig:R2vs1overR2}(b), but in
the form $1/\xi^2$ vs. $s$.  The data of cases 3 and 4 are well fitted
by a linear form
\begin{eqnarray}
\frac{1}{\xi^2}  = As + B 
\label{blobs_fit}
\end{eqnarray}
which supports the scaling form of \eqref{xi_blob}. In fitting the
data we used a non vanishing intercept $B$, due to the correction
from the last blob. In case 2 the last blob seems to be extended to
the whole polymer and the scaling $1/\xi^2 \sim s$ cannot be clearly
observed.  We have fitted the data of $1/\xi^2$ vs. $s$ in the range
of values where this relation is linear using a two-parameter fit
on \eqref{blobs_fit}. The slope coefficient $A$ as a function of
$\Omega$ is shown in the inset of \ref{fig:R2vs1overR2}(b). This
quantity increases linearly with $\Omega$ in agreement with the
prediction of \eqref{xi_blob}.  A similar analysis was performed for
self-avoiding polymers. The results are in agreement with the blob theory
and are shown in~\ref{fig:SAW_R2vs1overR2}(a,b). Close to the free
end the polymer exhibits equilibrium scaling $\xi^2\sim(N-s)^{2\nu}$
(\ref{fig:SAW_R2vs1overR2}(a)).  The plot $1/\xi^2$ vs. $s$
(\ref{fig:SAW_R2vs1overR2}(b)) shows the characteristic scaling behavior
$\sim 1/\sqrt{s}$ of torsional blobs \eqref{xi_blob}.

\begin{figure*}[b!]
\begin{center}
\includegraphics[width=0.45\textwidth]{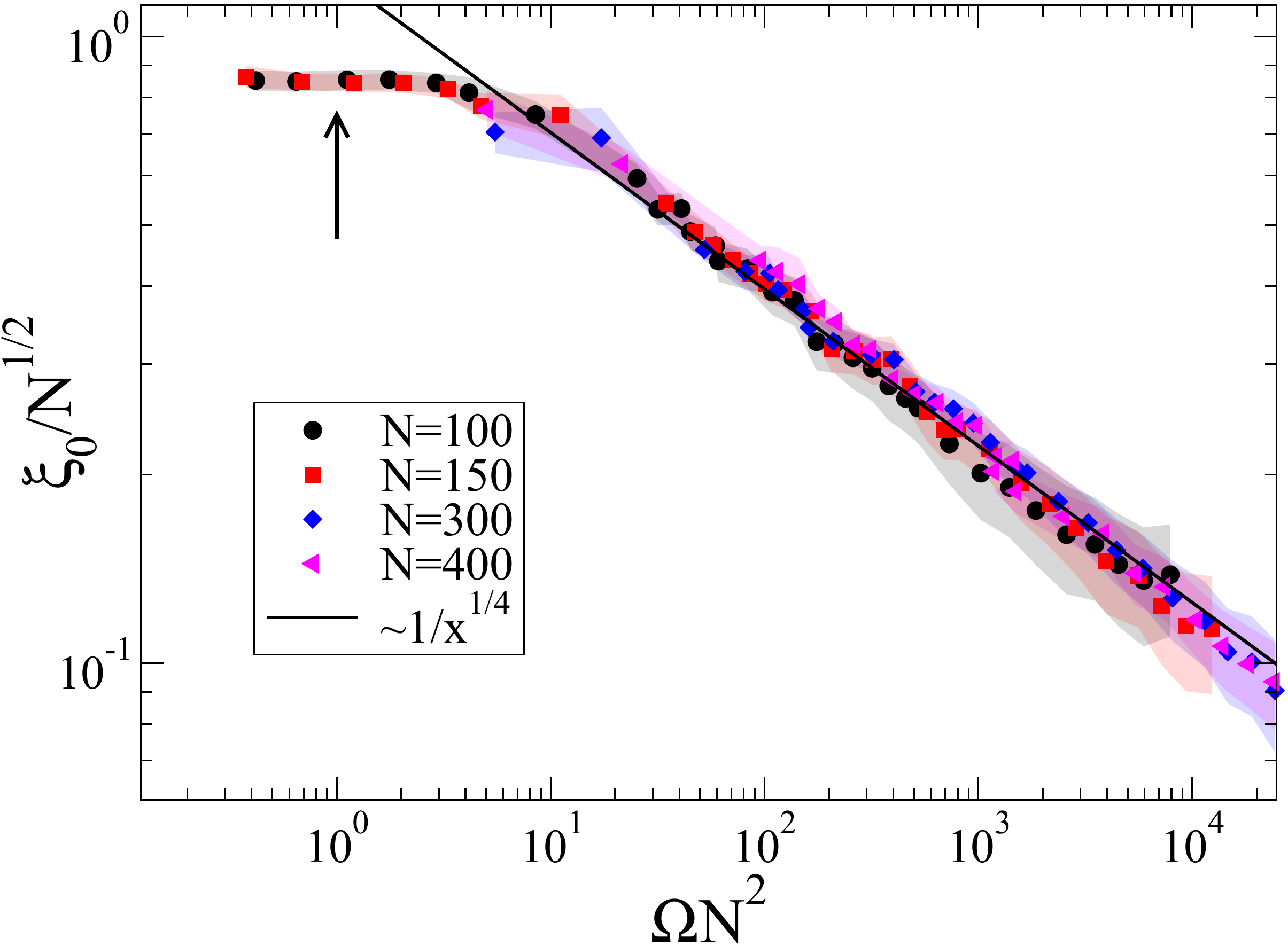}
\includegraphics[width=0.45\textwidth]{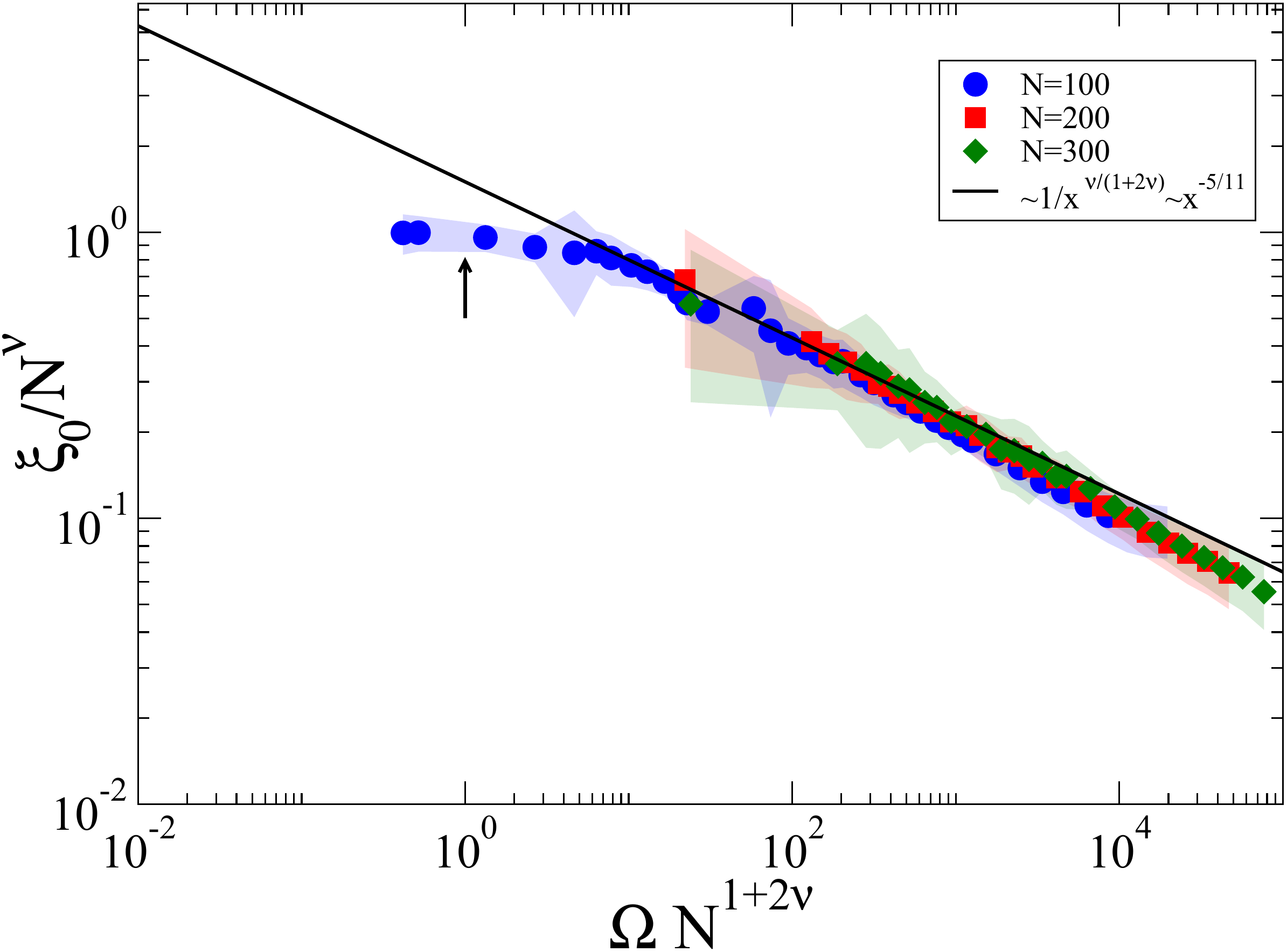}
\caption{Plot of $\xi_0/N^{1/2}$ vs. $\Omega N^{1+2\nu}$ for (a) an ideal
polymer ($\nu=1/2$) and (b) a self-avoiding polymer ($\nu=0.588$). The
data for different lengths collapse into a single master curve as
predicted from \eqref{scaling_xi0}. The scaling function governs the
crossover from equilibrium to trumpet regime, expected to take place for
$\tau_0 \Omega N^{1+2\nu} \simeq 1$, see \eqref{M_eq_reg}. This value
is indicated by a vertical arrow. Recall that in the simulation units
$\tau_0 \simeq 1$. 
The standard error is depicted as a shade around the plots.}
\label{fig_R2_vs_Omega}
\end{center}
\end{figure*}

\subsection{The last blob}

We focus now on the size of the last blob $\xi_0$.  In the trumpet regime,
according to \eqref{last_xi}, $\xi_0 \! \sim \! \Omega^{-\nu/(1 +
2 \nu)}$, while in the equilibrium case $\xi_0 \! \simeq \! a N^\nu$.
We can connect the two regimes using the following ansatz:
\begin{eqnarray}
\xi_0 = a N^\nu \, \phi\left( \tau_0 \Omega N^{1+2\nu} \right),
\label{scaling_xi0}
\end{eqnarray}
where $\phi(x)$ is a scaling function and the scaling variable
$x \! \equiv \! \tau_0 \Omega N^{1+2\nu}$, which is the natural combination
of $\Omega$ and $N$ (see \eqref{M_eq_reg}). The equilibrium regime
is recovered in the limit of small $x$ if $\phi(x) \to \phi_0 \simeq
1$. For large $x$ we require:
\begin{eqnarray}
\phi (x) \sim x^{-\frac{\nu}{1+2\nu}},
\label{phi_scal}
\end{eqnarray}
to recover \eqref{last_xi}. Note that the last blob is insensitive to
the transition from the trumpet to the stem-trumpet regime.

We determine the size of the last blob from simulation data using
\eqref{defxs} for $s=0$. \ref{fig_R2_vs_Omega} shows a plot of
$\xi_0/N^{0.5}$ vs. $\Omega N^{1+2\nu}$ for (a) ideal polymers ($\nu
=1/2$) and (b) self-avoiding polymers ($\nu =0.588$). The data for various
polymer lengths $N$ collapse into a single curve in agreement with the
scaling behavior predicted by the scaling ansatz \eqref{scaling_xi0}. For
large values of $\Omega N^{1+2\nu}$ the data follow the prediction of
\eqref{phi_scal}.  The arrows in \ref{fig_R2_vs_Omega} show the estimated
transition point from the equilibrium to the trumpet regime ($\tau_0
\Omega N^{1+2\nu}=1$, where we set $\tau_0=1$). Again the onset of the
trumpet regime occurs at somewhat higher values of $\Omega$ and $N$
than those predicted from the analytical arguments.

\begin{figure*}[t]
\begin{center}
\includegraphics[width=0.45\textwidth]{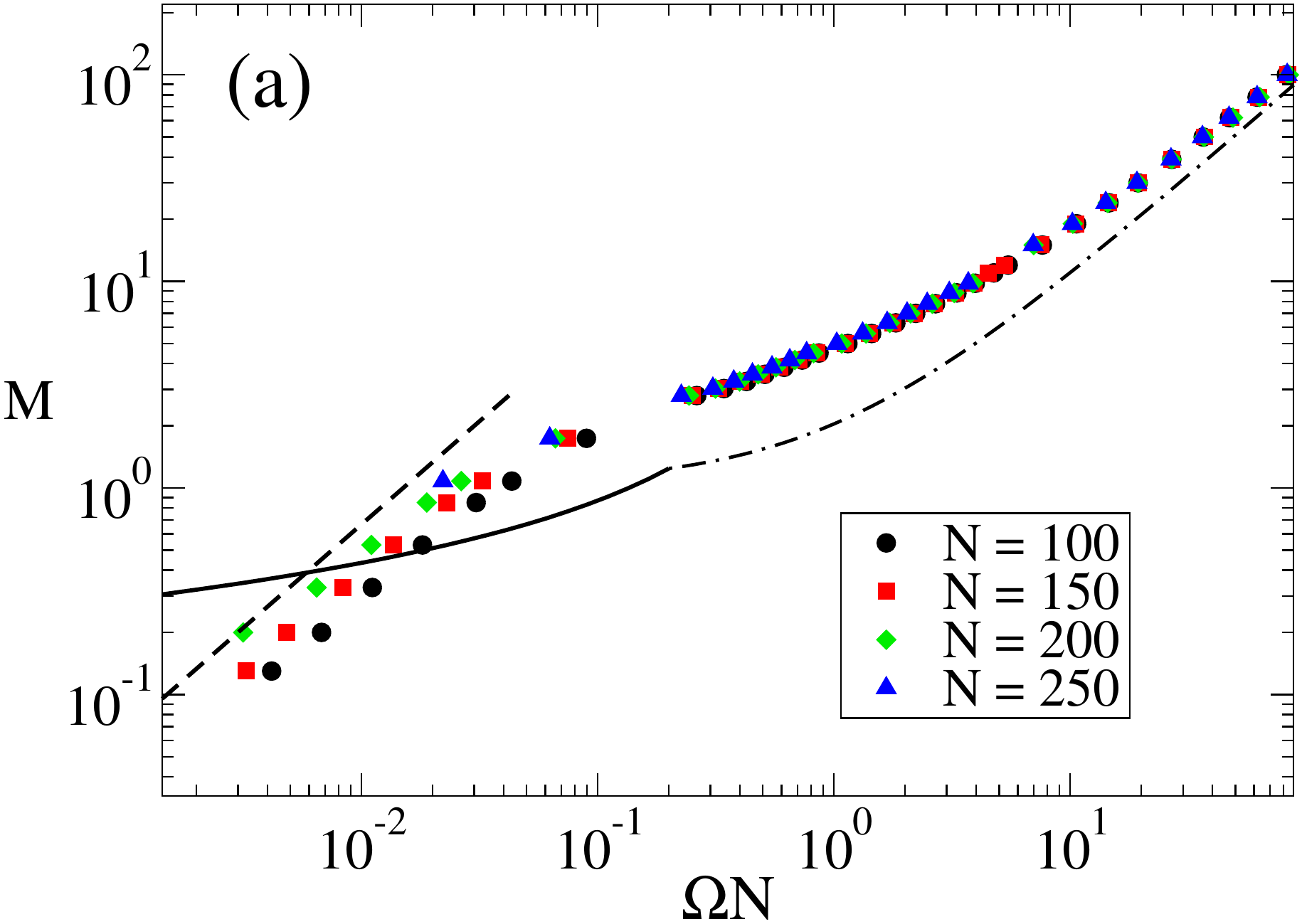}
\includegraphics[width=0.45\textwidth]{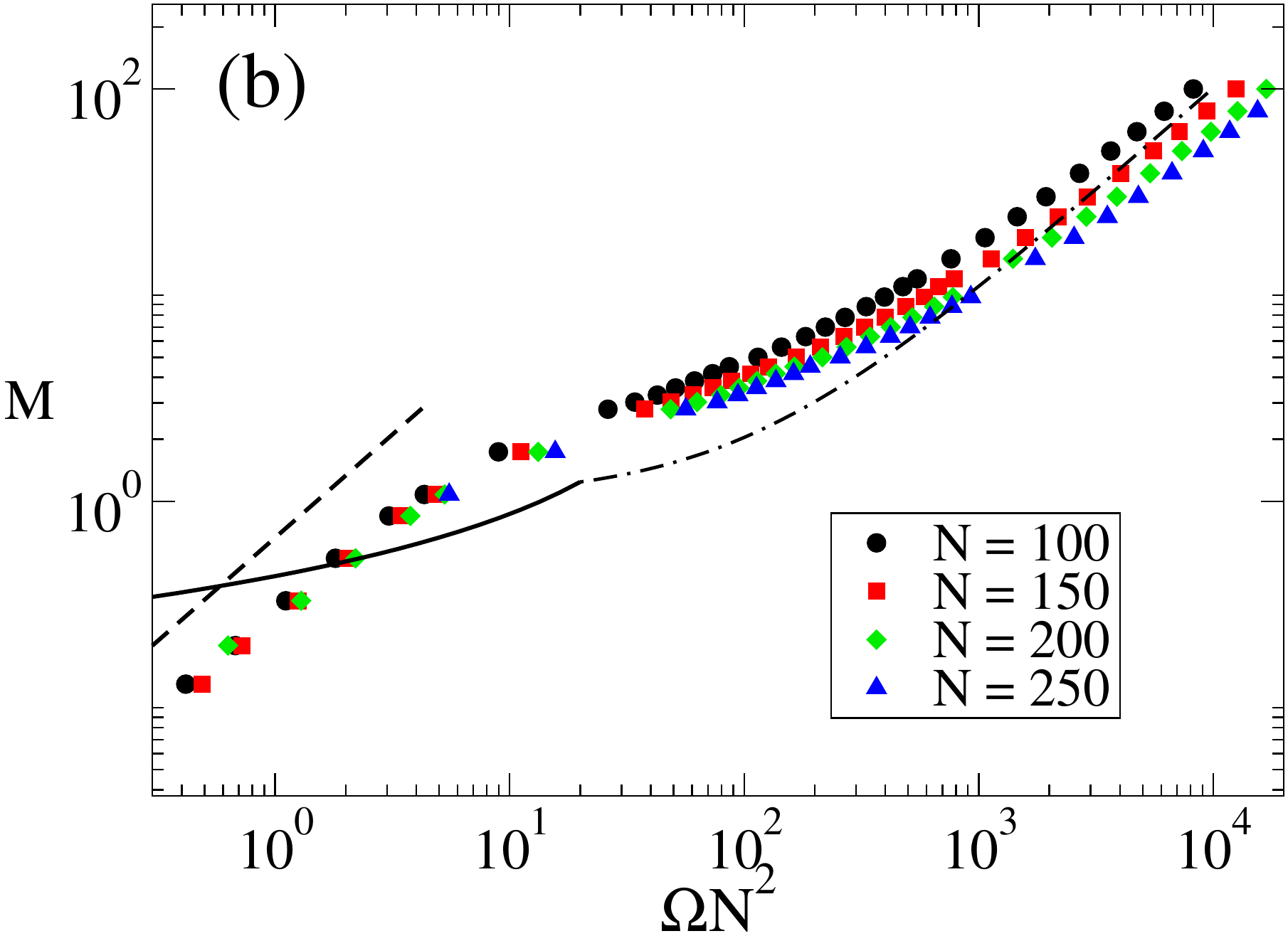}
\caption{Symbols: Plots of $M$ vs $\Omega N$ (a) and $\Omega N^{1+2\nu}$
(b) as obtained from simulations of ideal polymers ($\nu=1/2$) of
various lengths.  Dashed and solid lines: analytical predictions from
the blob theory (\eqref{M_eq}, \eqref{M_trump} and \eqref{M_stem})
for a polymer with length $N=100$.  These equations contain no adjustable
parameters. \eqref{M_eq} and \eqref{M_trump} are analytically
continued beyond their domain of validity, which produces a crossing of
the two lines.  In agreement with the blob theory, the simulation data
indicate that the torque is a function of the scaling variable $\Omega
N^{1+2\nu}$ in the equilibrium regime, while it is a function of $\Omega
N$ in the trumpet and stem-trumpet regimes. 
The standard error, which is in the horizontal direction since $\Omega$ 
was measured, is smaller than symbol size.}
\label{fig:MvsO}
\end{center}
\end{figure*}

\subsection{$\mathbf{M}$ vs. $\boldsymbol{\Omega}$}

\ref{fig:MvsO} shows a plot of the applied torque $M$ vs. $\Omega N$
(a) and vs. $\Omega N^{1+2\nu}$ (b) as obtained for the
simulations for polymers of various lengths (we consider here only the
case of ideal polymers ($\nu=1/2$)). The lines are the plots of the blob
theory predictions (\eqref{M_eq}, \eqref{M_trump} and \eqref{M_stem})
for a polymer of length $N=100$.  At low and high torques, the torque
is proportional to the angular velocity, which is in agreement with the
predictions of \eqref{M_eq} and \eqref{M_stem}. The two regimes are
separated by an intermediate (trumpet) regime, where ``thinning" occurs:
the torque increases more slowly than linear in $\Omega$. In the low
torque (equilibrium) regime the data from polymers of various lengths
collapse when plotted as function of $\Omega N^2$ (\ref{fig:MvsO}(b)),
in agreement with \eqref{M_eq}. At higher torques the data collapse
when plotted as functions of $\Omega N$ (\ref{fig:MvsO}(a)), which
is again in agreement with ~\eqref{M_trump} and \eqref{M_stem}.
There is an overall quite satisfactory quantitative agreement between
theory and simulations, keeping into account that we use no adjustable
parameters in \eqref{M_eq}, \eqref{M_trump} and \eqref{M_stem}
(we used $\tau_0 =1$ as estimated from our simulations units). Comparing
the simulations with the torsional blob theory predictions, we find that
the differences are at most within a factor three. We recall that the
torque-balance arguments in the blob theory are based on quasi-equalities
which ignore unknown numerical prefactors which are expected to be
of the order of unity.  A more detailed test of the dependence of $M$
on $\Omega$ in the trumpet regime (as predicted by \eqref{M_trump})
is at the moment not possible because logarithmic dependences are notoriously
difficult to establish as they require very long polymers.

\FloatBarrier

\begin{figure}[t]
\begin{center}
\includegraphics[width=0.8\textwidth]{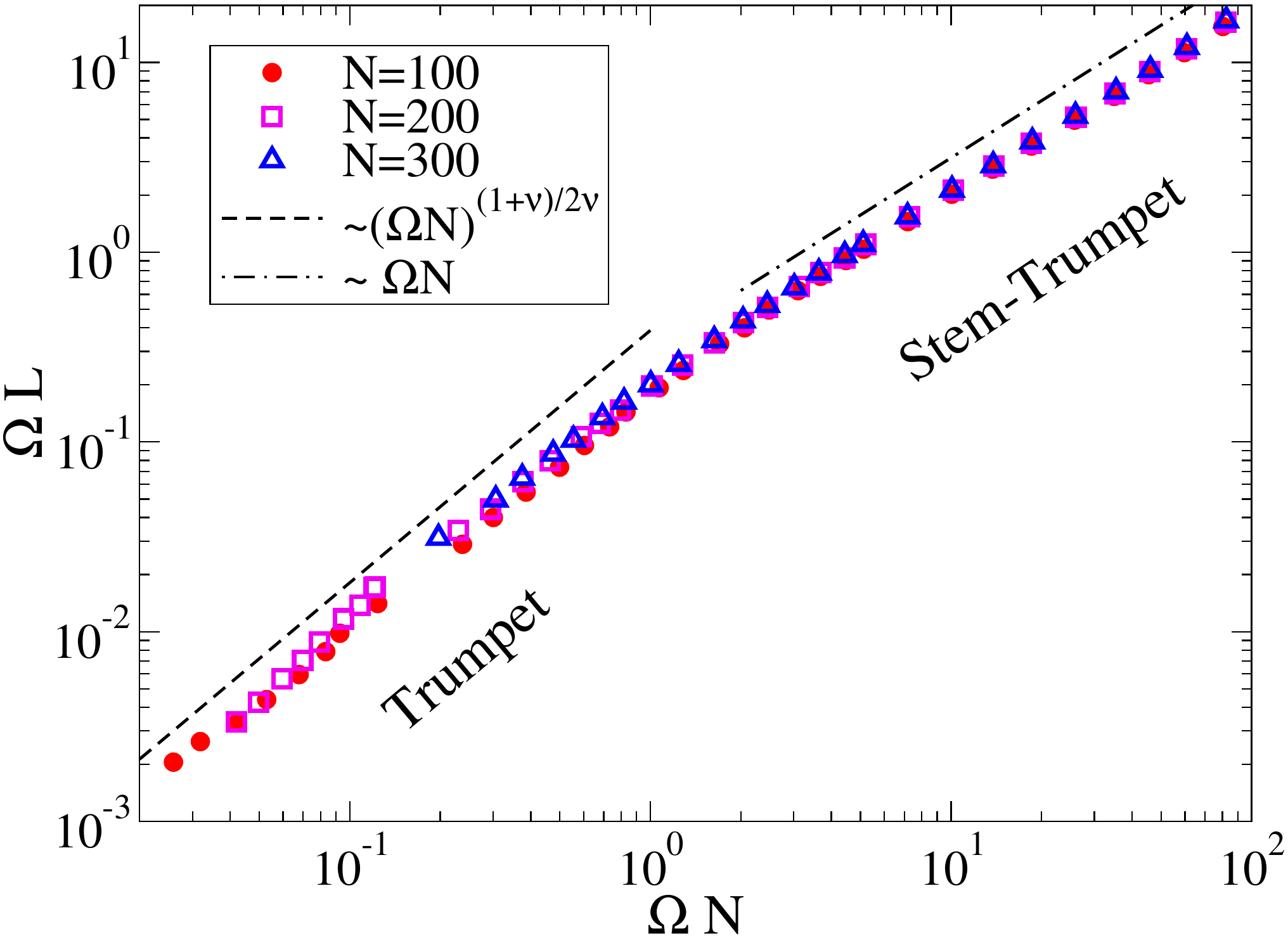}
\caption{Plot of $\Omega L$ vs. $\Omega N$ for three self-avoiding
polymers of different lengths.  The elongation $L$, defined in
\ref{def_elongation} is the distance along the rod axis between the
two end monomers. The simulation data agree well with the blob theory,
which predicts two different scaling behaviors in the trumpet $\Omega L
\sim (\Omega N)^{1.33}$ and in the stem-trumpet $\Omega L \sim \Omega N$
regimes (the Flory exponent for self-avoiding polymers is $\nu \approx
0.588$). The standard error is smaller than symbol size.}
\label{fig:elongation}
\end{center}
\end{figure}

\subsection{Elongation along the rod}

The elongation of the polymer along the rod was computed only for the 
case of self avoidance where \cite{degennes} $\nu \approx 0.588$. 
In the simulation
we determined the average absolute distance between the $z$-coordinates
of the first and last monomer, so the elongation is defined as
\begin{eqnarray}
L = \left\langle \left| z(0) - z(N)\right| \right\rangle.
\label{def_elongation}
\end{eqnarray}
As we are interested in the behavior of $L$ in the trumpet and
stem-trumpet regimes we plot in \ref{fig:elongation} $\Omega L$
vs. $\Omega N$ for polymers of various lengths.  For self-avoiding
polymers the theoretical prediction \eqref{Ltrumpet} implies $\tau_0
\Omega L \simeq a (\tau_0 \Omega N )^{1.33}$.  In the stem trumpet regime,
\eqref{Lstem} yields to leading order in $N$: $\tau_0 \Omega L \simeq
h \tau_0 \Omega N$.  The crossover between these two regimes around $
\tau_0 \Omega N \simeq 1$ \eqref{onset_stem} is indeed observed in
\ref{fig:elongation}, where the data belonging to polymers of different
lengths collapse into a single curve.

\section{Conclusions}

In this paper, we gave a full description of the {\it stationary}
rotation of a flexible polymer induced by a torque.  Using a local
torque-balance equation, we computed the polymer shape as a function
of the curvilinear coordinate $s$. We identify three regimes as the
angular velocity increases (or equivalently, when the torque applied to
the monomer attached to the rod increases).  (i)~The equilibrium regime,
where the polymer is not perturbed from its equilibrium configuration.
(ii)~The trumpet regime, in which the polymer is composed of blobs of
decreasing size, starting from the free end. Here the blob size decays
as $1/\sqrt s$, which is much slower than the case of a constant force
applied at one end ($\sim 1/s$). Finally (iii) the stem-trumpet regime
is defined when a part of the polymer close to the attached monomer
is tightly wound around the axis while the free end still displays a
trumpet shape.

One of the central results of this paper is the characterization
of the response of the total torque $M$ to a change of the angular
velocity $\Omega$ or the number of monomers $N$. In the equilibrium
and the stem-trumpet regime, the torque is proportional to the
angular velocity $M \simeq \Gamma \Omega$.  The torque friction,
$\Gamma$, has a different scaling behavior with the polymer length in
equilibrium $\Gamma\propto N^{1+2\nu}$ compared to the stem-trumpet
regime $\Gamma\propto N$.  More surprisingly, in the trumpet regime
the torque has a very weak (logarithmic) dependence on $\Omega$. This
dependence stems from the fact that when the torque is increased the
polymer gets closer to the central axis hereby reducing the resistance
to the rotational motion.  Non-power law, and specifically logarithmic,
behavior has already been predicted, simulated and observed in the case
of polymer stretching.\cite{saleh2009, stevens2013} However, in that
case a situation where both ends are subjected to a force is studied
and the logarithmic dependence represents an increase of resistance for
high forces due to short-scale DNA crumpling.  The setup we consider is
different: the resistance decreases in an intermediate torque regime,
due to the decrease in distance between the polymer and the rod.

The torque-torsional blob relation \eqref{scal_M_xi} is an analogue
of the well-known force-tensile blob relation in the pulling problem
\cite{pincus76}, and is expected to play an essential role to
further explore the rotational dynamics. It would be interesting to
check experimentally the predictions of the theory. The experimental
protocol could be achieved with excitable cylinders in an optical torque
wrench~\cite{pedaci11}. A polymer could be attached to the cylinder
and the other end of the polymer labelled with a fluorescent molecule,
allowing, e.g., to test the dependence of spacing of the free end from
the axis of rotation with respect to the angular velocity.

\section{Acknowledgement}

B.B. acknowledges the support of  Grant No CA077712 from the National
Cancer Institute, NIH to the Hanawalt laboratory at Stanford University.
T.S. is supported by KAKENHI [Grant No.26103525, ``Fluctuation and
Structure", Grant No.24340100, Grant-in-Aid for Scientific Research
(B)], Ministry of Education, Culture, Sports, Science and Technology
(MEXT), Japan and JSPS Core-to-Core Program (Nonequilibrium Dynamics of
Soft Matter and Information).  J-C.W. acknowledges the support by the
Laboratory of Excellence Initiative (Labex) NUMEV, OD by the Scientific
Council of the University of Montpellier.


\providecommand*{\mcitethebibliography}{\thebibliography}
\csname @ifundefined\endcsname{endmcitethebibliography}
{\let\endmcitethebibliography\endthebibliography}{}

\clearpage


\begin{figure}[t]
\begin{center}
\includegraphics[width=0.65\textwidth]{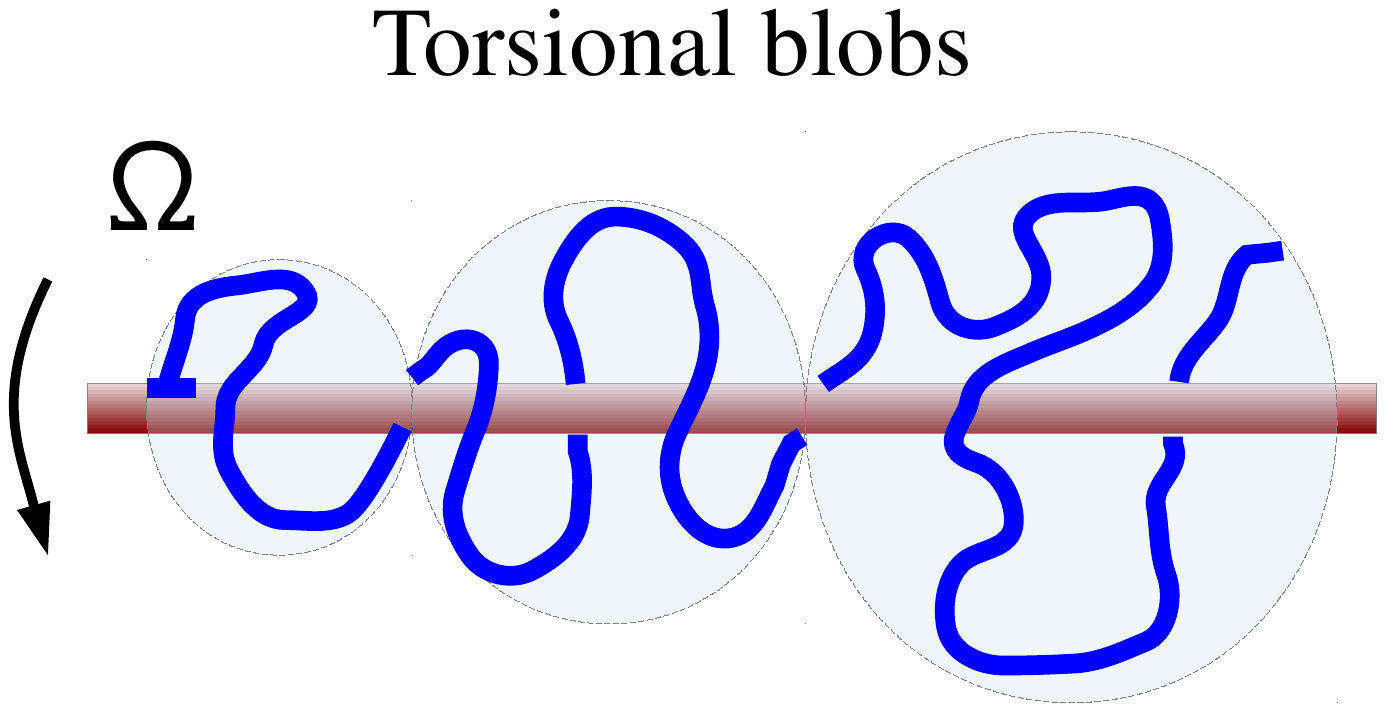}
\end{center}
\end{figure}

\section*{Table of Contents graphic}

\subsection*{Torque-induced rotational dynamics in polymers: Torsional blobs and thinning}

{Michiel Laleman}, {Marco Baiesi}, {Boris P. Belotserkovskii}, {Takahiro
Sakaue}, {Jean-Charles Walter} and {Enrico Carlon}

\end{document}